\documentclass[12pt,amsart,superscriptaddress,floatfix]{iopart}
\usepackage{fullpage,graphicx,epsf}
\usepackage{iopams}%
\usepackage{setstack}
\usepackage{amsfonts,amssymb}
\usepackage{amscd}
\usepackage{cite}
\usepackage{color}
\usepackage{graphicx}
\usepackage{epsfig}
\usepackage[english]{babel}
\usepackage{amsfonts}
\usepackage{soul}
\begin{document}
\def\l{\label}
\def\p{\partial}
\def\be{\begin{equation}}
\def\ee{\end{equation}}
\def\fr{\frac}
\def\epsilon{\varepsilon}
%=============================================================================
\title{Kinetic theory of nonequilibrium stochastic long-range systems:
Phase transition and bistability}
\author{Cesare Nardini$^{1,2}$, Shamik Gupta$^1$, Stefano Ruffo$^{1,3}$,
Thierry Dauxois$^{1}$ and Freddy Bouchet$^1$}
\address{$^1$ Laboratoire de Physique de l'Ecole Normale Sup\'erieure de Lyon, Universit\'e de Lyon and CNRS, 46, All\'ee d'Italie, F-69007 Lyon, France}
\address{$^2$ Dipartimento di Fisica e Astronomia and CSDC, Universit\`a di Firenze and INFN, via G. Sansone 1, I-50019 Sesto Fiorentino (FI), Italy}
\address{$^3$ Dipartimento di Energetica ``Sergio Stecco" and CSDC,
Universit\`a di Firenze, CNISM and INFN, via S. Marta 3, 50139 Firenze, Italy}
\ead{cesare.nardini@gmail.com,shamikg1@gmail.com,\\stefano.ruffo@gmail.com,thierry.dauxois@ens-lyon.fr,freddy.bouchet@ens-lyon.fr}

\begin{abstract}
We study long-range interacting systems driven by external stochastic
forces that act collectively on all the particles constituting the
system. Such a scenario is frequently encountered in the context of
plasmas, self-gravitating systems, two-dimensional turbulence, and
also in a broad class of other systems. Under the effect of stochastic
driving, the system reaches a stationary state where
external forces balance dissipation on average. These states have the
invariant probability that does not respect detailed balance, and are characterized by non-vanishing
currents of conserved quantities. In order to
analyze spatially homogeneous stationary states, we develop a kinetic approach that generalizes the one known for deterministic long-range
systems; we
obtain a very good agreement between predictions from kinetic theory and
extensive numerical simulations. Our approach may also be generalized
to describe spatially inhomogeneous stationary states. We also report on
numerical simulations exhibiting a
first-order nonequilibrium phase transition from homogeneous to
inhomogeneous states. Close to the phase transition, the system shows bistable behavior between
the two states, with a mean residence time that 
diverges as an exponential in the inverse of the strength of the external stochastic
forces, in the limit of low values of such forces.  
\end{abstract}
\date{\today}
\maketitle
%=============================================================================
\section{Introduction}
Systems of particles interacting through
two-body non-integrable potentials, also called long-range
interactions, abound in nature. Common examples are plasmas interacting
through repulsive or attractive Coulomb potential, self-gravitating
systems (globular clusters, galaxies) involving interaction through attractive Newton
potential, two-dimensional turbulence, and many others. In addition, there are several model systems
with non-integrable interactions which have been studied extensively in
recent years, such as spins, vortices in two dimensions, etc \cite{Campa:2009,Bouchet:2010,Chavanis,JStatMech,Bouchet-Venaille}.
 
Very often, these systems are acted upon by external stochastic forces
that drive them out of equilibrium. Unlike systems with short-range
interactions, stochastic forces in long-range interacting systems act coherently on all particles, and not
independently on each particle. Consider,
e.g., globular clusters being influenced by the
gravitational potential of their galaxy, which produces a force that
fluctuates along their physical trajectories. 
In addition, galaxies themselves feel the random potential of other
surrounding galaxies, and their halos are subjected to transient and
periodic perturbations, which may be due to the passing of dwarfs or to
orbital decaying~\cite{Weinberg2000}. Dynamics of plasmas are also
strongly influenced by fluctuating electric and magnetic fields due to
the ever-changing ambiance \cite{Liewer}. In situations of stochastic
driving, the systems at long times often reach a nonequilibrium
stationary state that violates detailed balance. In such a state, the power injected by the external random fields balances on average the
dissipation, and there is a steady flux of conserved quantities through
the system.

Study of nonequilibrium stationary
states (NESS) is an active area of research of modern day statistical
mechanics. One of the primary challenges in this field is to formulate a
tractable framework to analyze nonequilibrium systems on a
common footing, similar to the one due to Gibbs and Boltzmann that has
been established for equilibrium systems~\cite{Derrida:2007,Jarzynski:2008,Dhar:2008}.
This paper provides, to our knowledge, the first study of NESS in
long-range systems with statistical mechanical perspectives.

Common theoretical approaches to study isolated systems with
long-range interactions include the kinetic theory description of 
relaxation towards equilibrium. In plasma physics and astrophysics, this approach leads
to the Lenard-Balescu equation or to the approximate Landau equation~\cite{Nicholson:1992,Lifshitz:2002}.
One of the main theoretical results of this paper is a detailed
development of a generalization of this kinetic theory approach to
describe nonequilibrium stationary states in systems with long-range
interactions driven by external stochastic
forces, valid in the limit of small
external stochastic fields. 
The nonequilibrium kinetic equation that we obtain describes the temporal evolution of
the one-particle distribution function. In the limit of small external
forcing, the system settles into a
stationary state, in which we find the one-particle
momentum distribution to be non-Gaussian. The predictions of our kinetic
equation for spatially homogeneous stationary states compare very well with
results of our extensive $N$-particle numerical simulations on a paradigmatic model of 
long-range interacting systems. Our numerical simulations also exhibit a
nonequilibrium phase transition between homogeneous and inhomogeneous
states. Close to the phase transition, we demonstrate the
occurrence of bistability between these two types of states, with a mean
residence time that
diverges as an exponential in the inverse of the strength of the external forcing, in the
limit of low values of such forcing.

Similar bistable behavior has recently been observed in two-dimensional
turbulence with stochastic forcing \cite{Bouchet-Simonnet}. We believe
that such phase transitions are essential phenomena for geophysical
flows and climate, for which the two-dimensional Euler equations are a
simplified paradigmatic model.  There exists a very strong analogy
between the two-dimensional Euler equations and the Vlasov equation
relevant for leading order dynamics of the model we discuss in this
paper \cite{Briggs,Dikii}. One of the motivations of the present work is to be able to study analogous phenomena in a setup for which the theory can be more easily worked out.

In a recent letter, we reported on some of the above findings, in one of the first studies of non-equilibrium stationary
states in systems with non-integrable potentials and driven by external stochastic fields \cite{Nardini:2012}.  The aim of this paper is to present a detailed derivation of the results given in Ref. \cite{Nardini:2012}, as well as to report on additional empirical results, more specifically non-equilibrium phase transitions.

We note that  Ref. \cite{Chavanis:2012} presents a computation of the
same kinetic equation as the one we describe in this work and in
\cite{Nardini:2012}. Nevertheless, the result is different.

The structure of the paper is as follows. In the following section, we
define the dynamics we are going to consider of a long-range interacting
system driven by external stochastic forces. We also discuss the
paradigmatic example of the Hamiltonian mean-field (HMF) model.
In section \ref{methods}, we discuss the methods we adopt to analyze the
dynamics. In particular, we give a detailed derivation of the kinetic
theory to study spatially homogeneous stationary states of the dynamics.
We describe the numerical simulation scheme that we employ to study
the dynamics, specifically, to check the predictions of our kinetic
theory. This is followed by a discussion in section \ref{results1} of the
results obtained from the kinetic theory, and their 
comparison with numerical simulation results for spatially homogeneous stationary
states. In section ~\ref{results2}, we discuss the results of numerical simulations of
spatially inhomogeneous stationary states. We report on the very
interesting bistable behavior in which the system in the course of its
temporal evolution switches back and forth between homogeneous and
inhomogeneous states, with a mean residence time that we show to be diverging
as an exponential in the inverse of the strength of the external
stochastic forcing, in the limit of low values of such forcing.
We close the paper with concluding remarks. Some of the technical
details of our computation are collected in the four appendices. 
%=============================================================================
\section{Long-range interacting systems driven by stochastic fields}
\label{model}
\subsection{The model}
Consider a system of $N$ particles interacting through a long-range
pair potential, and described by the Hamiltonian
\begin{equation}\label{Hamiltonian}
H=\sum_{i=1}^N\frac{p_i ^2}{2}+\frac{1}{2N}\sum_{i,j=1}^N v(q_i-q_j).
\end{equation}
Here, $q_i$ and $p_i$ are, respectively, the coordinate and the momentum
of the $i$-th particle, and $v(q)$ is the two-body interaction
potential. We take the particles to be of unit mass. In this paper, for
simplicity, we regard $q_i$'s as scalar periodic variables of period $2\pi$; generalization to $q_i \in
\mathbb{R}^n$, with $n=1, 2$ or $3$, is straightforward.  

In plasma physics, the typical  number of
particles interacting with one particle is given by the coupling
parameter $\Gamma = n\lambda_D^3$, where $n$ is the number density, and
$\lambda_D$ is the Debye length. It is then usual to rescale time
such that the inverse of $\Gamma$ multiplies the interaction
term~\cite{Nicholson:1992}. In self-gravitating systems, the dynamics is dominated by
collective effects, so that it is natural to rescale time in such a
way that the parameter $1/N$ multiplies the interaction
potential~\cite{Milion-body-pb}. These facts explain the rescaling of the
potential energy by $1/N$ in Eq.~(\ref{Hamiltonian}), called the
Kac scaling in systems with long-range interactions~\cite{Kac}. We
emphasize that no generality is lost in adopting the Kac prescription.

We perturb the system~(\ref{Hamiltonian}) by a statistically homogeneous Gaussian stochastic field $F(q,t)$ with zero mean, and variance given by 
\begin{equation}
\langle F(q,t)F(q',t')\rangle = C(|q-q'|)\delta(t-t').
\end{equation}
The resulting equations of
motion for the $i$-th particle are
\begin{eqnarray}
\label{equations_of_motion}
\dot{q}_i=\frac{\p H}{\p p_i},\qquad {\rm and} \qquad
\dot{p}_i=-\frac{\p H}{\p q_i}-\alpha p_i +\sqrt{\alpha}\,F(q_i,t).
\l{dynamics}
\end{eqnarray}

The property that the Gaussian fields $F(q_i,t)$ are statistically homogeneous,
i.e., the correlation function $C$ depends solely on $|q-q'|$, is
consistent with any perturbation that respects space homogeneity. Such a
property is necessary for the discussions later in the paper on the
kinetic theory approach to describe spatially homogeneous stationary states of the
dynamics~(\ref{equations_of_motion}). Note that $C(q)$ is the
correlation, so that it is a
positive-definite function~\cite{Papoulis}, and its Fourier components are positive:
\begin{equation}\label{eq:g}
c_k \equiv \frac{1}{2\pi}\int_0^{2\pi}{\rm d}q ~C(q)e^{-ikq} > 0;
~~c_{-k}=c_k,~~~~ C(q)=c_0+2\sum_{k=1}^{\infty}c_k \cos(kq).
\label{ckdefinition}
\end{equation}
We find it convenient to use the equivalent Fourier representation of
the Gaussian field $F(q,t)$ as follows:
\begin{eqnarray}\label{stoc_forces} 
F(q,t)\,=\,\sqrt{c_0}\ X_0(t) + \sum_{k=1}^{\infty} \sqrt{2
c_k}\left[\cos (kq) \,X_k(t) + \sin
(kq) \, Y_k(t)\right],
\label{noise-form}
\end{eqnarray}
where $X_0(t)$, $X_k(t)$ and $Y_k(t)$ are independent scalar Gaussian white noises satisfying
\begin{eqnarray}
&&\langle X_k(t)\,X_{k'}(t')\rangle=\delta_{k,k'}\delta(t-t'), \\
&&\langle Y_k(t)\,Y_{k'}(t')\rangle=\delta_{k,k'}\delta(t-t'), \\
&&\langle X_k(t)\,Y_{k'}(t')\rangle=0.
\end{eqnarray}
For stochastic dynamics of the type Eq. (\ref{equations_of_motion}),
general mathematical results allow to prove that the dynamics is ergodic
(see, for instance, \cite{Bellet}).

Using the It\={o} formula~\cite{Gardiner} to compute the time derivative of
the energy density $e=H/N$, and averaging over noise realizations
give
\begin{equation}\label{evolution_H} 
\left\langle\frac{de}{dt}\right\rangle+\left \langle 2\alpha \kappa\right\rangle=\frac{\alpha}{2}C(0),
\label{kinetic-energy}
\end{equation}
where $\kappa=\sum_{i=1}^Np_i ^2/(2N)$ is the kinetic energy density. On
integration, we get for homogeneous states for which $e=\kappa$ that 
\begin{equation}
\langle k(t) \rangle \,=\,\left(\langle k(0)\rangle-\frac{C(0)}{4}\right)\,e^{-2\alpha t}\,+\,\frac{C(0)}{4}.
\label{kinetic-energy-1}
\end{equation}
The average kinetic energy density in the stationary state is thus
$\left\langle \kappa\right\rangle_{ss} = C(0)/4$. We define the kinetic
temperature of the system as
\begin{equation}
\left\langle \kappa\right\rangle_{ss}\equiv \frac{T}{2};
\label{kinetic-energy-steady-state}
\end{equation}
as a result, we have
\begin{equation} 
T=\frac{C(0)}{2}.
\label{kinetic-temp-defn}
\end{equation}

Let us note that in the dynamics~(\ref{equations_of_motion}), fluctuations of intensive observables due
to stochastic forcing are of order $\sqrt{\alpha}$, while those due to
finite-size effects are of order $1/\sqrt{N}$. Moreover, the typical timescale
associated with the effect of stochastic forces is $1/\alpha$, as is
evident from Eq.~(\ref{kinetic-energy-1}), while the one associated with
relaxation to equilibrium due to finite-size effects is of order $N$,
see~\cite{Campa:2009,Bouchet:2010}.   

Our theoretical analysis to study the dynamics (\ref{equations_of_motion}) by means of kinetic theory is valid for
any general two-particle interaction potential $v(q)$. However, in order to perform simple numerical
simulations with which we may check the predictions of the kinetic
theory, we specifically make the choice $v(q)=1-\cos q$, that defines
the stochastically-forced attractive Hamiltonian mean-field (HMF) model,
as detailed below.
\subsection{A specific example: The stochastically-forced HMF model}
\label{HMFmodel}
The Hamiltonian mean-field (HMF) model is a paradigmatic model to study long-range
interacting systems. The model describes particles moving on a circle under
deterministic Hamiltonian dynamics, and interacting through the
interparticle potential $v(q)=1-\cos q$ \cite{yamaguchi2004,Antoni}. It displays many features of
generic long-range interacting systems, e.g., existence of
quasistationary states~\cite{yamaguchi2004,Campa:2009}. In equilibrium, the
model has a second-order phase transition from a high-energy spatially homogeneous phase
to a low-energy inhomogeneous phase at the energy density $e_c=3/4$,
corresponding to the critical temperature $T_c=1/2$. In
a system of $N$ particles, the
degree of spatial inhomogeneity at time $t$ is measured by
the magnetization variable $m(t)$, defined as
\begin{equation}
m(t)=\frac{1}{N}\sqrt{\Big(\sum_{i=1}^N\cos
q_i\Big)^2+\Big(\sum_{i=1}^N\sin q_i\Big)^2}.
\l{magnetization}
\end{equation}
In the thermodynamic limit $N \to \infty$, the magnetization in the
steady state decreases continuously as a function of energy from one to
zero at the transition energy $e_c$, and remains zero at all higher
energies. When forced by the stochastic forces $F(q_i,t)$ resulting in the
dynamics (\ref{equations_of_motion}), we call the corresponding model the stochastically-forced HMF model.
We note for later purpose that the Fourier transform of the HMF interparticle potential is, for $k \neq 0$,
$v_k=-\left[\delta_{k,1}+\delta_{k,-1}\right]/2$, where $\delta_{k,i}$
is the Kronecker delta.
%=============================================================================
\section{Methods of analysis}
\label{methods}
\subsection{\label{sec:Kinetic-theory}Kinetic theory for homogeneous
stationary states}
Here, we develop a suitable kinetic theory description to
study the dynamics
(\ref{equations_of_motion})
in the joint limit $N \to \infty$ and $\alpha \to 0$. While the first
limit is physically motivated on grounds that most long-range systems
indeed contain a large number of particles, the second one allows
us to study stationary states for small external forcing. 
Moreover, for small $\alpha$, we will be able to develop a complete kinetic theory for the dynamics.
For simplicity, we discuss here the continuum limit $N\alpha \gg 1$,
when stochastic effects are predominant with respect to finite-size
effects. The generalization of the following discussion to the cases
$N\alpha$ of order one and $N\alpha \ll 1$ is straightforward, as pointed
out at the end of this subsection. For the development of the kinetic
theory, we assume the
system to be spatially homogeneous; a possible generalization to the non-homogeneous case will be discussed in the conclusions of the paper.

As a starting point to develop the theory, we consider the Fokker-Planck
equation associated with the equations of motion
(\ref{equations_of_motion}). 
This equation describes the evolution of the $N$-particle distribution function
$f_{N}(q_{1},...,q_{N},p_{1},...,p_{N},t)$, which is the probability
density (after averaging over noise realizations) to observe
the system with coordinates and momenta around the values $\{q_{i},p_{i}\}_{1\leq i\leq N}$
at time $t$. This equation can be derived by standard methods
\cite{Gardiner}; we have
\begin{eqnarray}
\frac{\partial f_{N}}{\partial
t}&=&\sum_{i=1}^{N}\left[-p_{i}\frac{\partial f_{N}}{\partial
q_{i}}+\frac{\partial(\alpha p_{i}f_{N})}{\partial p_{i}}\right]+\frac{1}{2N}\sum_{i,j=1}^{N}\frac{\partial v(q_{i}-q_{j})}{\partial
q_{i}}\left[\frac{\partial}{\partial p_{i}}-\frac{\partial}{\partial
p_{j}}\right]f_{N} \nonumber \\
&&+\frac{\alpha}{2}\sum_{i,j=1}^{N}C(|q_{i}-q_{j}|)\frac{\partial^{2}f_{N}}{\partial p_{i}\partial p_{j}}.
\label{eq:N-particle-Fokker-Planck}
\end{eqnarray}

In \ref{detailed-balance-proof}, by applying the
so-called potential conditions \cite{Risken} for the above Fokker-Planck
equation, we prove that a necessary and sufficient condition for the
stochastic process~(\ref{equations_of_motion}) to verify detailed balance is that the
Gaussian noise is white in space, that is, $c_k=c$ for all $k$. This
condition is not satisfied for a generic correlation function $C(q)$, in
which case, the steady states of the dynamics are true nonequilibrium
ones, characterized by non-vanishing probability currents in
configuration space, and a balance between external forces and dissipation.

Similar to the Liouville equation for Hamiltonian systems, the $N$-particle
Fokker-Planck equation~(\ref{eq:N-particle-Fokker-Planck})
 is a very detailed description of the system.
Using kinetic theory, we want to describe the evolution of the one-particle
distribution function
\begin{equation}
f(z_{1},t)=\int\prod_{i=2}^{N}\,{\rm d}z_{i}f_{N}(z_{1},...,z_{N},t),
\end{equation}
where we have used the notation $z_{i} \equiv (q_{i},p_{i})$.
We note that with this definition, the normalization is $\int{\rm d}zf(z,t)=1$.
Substituting in the Fokker-Planck equation (\ref{eq:N-particle-Fokker-Planck})
the reduced distribution functions
\begin{equation}
f_{s}(z_{1},...,z_{s},t)=\frac{N!}{(N-s)!N^s}\int\prod_{i=s+1}^{N}\,{\rm d}z_{i}\, f_{N}(z_{1},...,z_{N},t),
\end{equation}
and using standard techniques \cite{Huang}, we get a hierarchy of
equations, similar to those of the Bogoliubov-Born-Green-Kirkwood-Yvon
(BBGKY) hierarchy, as follows:
\begin{eqnarray}\label{BBGKY-stochastic}
&&\frac{\partial f_{s}}{\partial t}+\sum_{i=1}^{s}p_{i}\frac{\partial
f_{s}}{\partial q_{i}}-\frac{1}{N}\sum_{i,j=1}^{s}\frac{\partial
v(q_{i}-q_{j})}{\partial q_{i}}\frac{\partial f_{s}}{\partial
p_{i}}-\sum_{i=1}^{s}\frac{\partial}{\partial p_i}[\alpha
p_{i}f_{s}]\nonumber \\
&&-\frac{\alpha}{2}\sum_{i,j=1}^{s}C(|q_{i}-q_{j}|)\frac{\partial^{2}f_{s}}{\partial
p_{i}\partial p_{j}}=\sum_{i=1}^{s}\int dz_{s+1}\frac{\partial}{\partial q_{i}}v(q_{i}-q_{s+1})\frac{\partial f_{s+1}}{\partial p_{i}}
\end{eqnarray}
for $s=1,...,N-1$. In this paper, we use both the
notations $\partial h/\partial q$ and $h'$ to denote the derivative of a
function $h$. With a slight
abuse of the standard vocabulary, we will refer to Eq.~(\ref{BBGKY-stochastic}) as the BBGKY hierarchy equation.

Now, as is usual in kinetic theory, we split the reduced distribution functions into connected and non-connected
parts, e.g., 
\begin{equation}
f_{2}(z_{1},z_{2},t)=f(z_{1},t)f(z_{2},t)+\tilde{g}(z_{1},z_{2},t),
\end{equation}
and similarly, for other $f_{s}$'s with $s >2$. In \ref{BBGKYclosure},
we show that the connected part $\tilde{g}(z_{1},z_{2},t)$ of the
two-particle correlation is of order~$\alpha$, so that we
may write
\begin{equation}
f_{2}(z_{1},z_{2},t)=f(z_{1},t)f(z_{2},t)+\alpha g(z_{1},z_{2},t),
\end{equation}
where $g$ is of order unity; more generally, the connected part of the $k$-particle correlation is 
of higher order, with respect to $\alpha$, in the small parameters $\alpha$ and $1/N$. Then, to close the BBGKY hierarchy, we neglect the effect of the connected
part of the three-particle correlation on the evolution of the
two-particle correlation function. This scheme is justified at leading order in
the small parameter $\alpha$, and is the simplest self-consistent closure
scheme for the hierarchy while taking into account the effects of the
stochastic forcing. With our assumption that
the system is homogeneous, i.e., $f$ depends on $p$, and $g$ depends
on $|q_{1}-q_{2}|$, $p_{1}$ and $p_{2}$ only, the first two equations
of the hierarchy are then 
\begin{equation}\label{eq:stochastic-BBGKY-1-eq}
\frac{\partial f}{\partial t}-\alpha\frac{\partial}{\partial
p}[pf]-\frac{\alpha}{2}C(0)\frac{\partial^{2}f}{\partial
p^{2}}=\alpha\frac{\partial}{\partial p}\int{\rm d}q_{1}{\rm
d}p_{1}v'(q_{1})g(q_{1},p,p_{1},t),
\end{equation}
and
\begin{eqnarray}\label{eq:stochastic-BBGKY-2-eq}
\frac{\partial g}{\partial t}+L_{f}^{(1)}g+L_{f}^{(2)}g & = &
C(|q_{1}-q_{2}|)f'(p_{1},t)f'(p_{2},t),
\end{eqnarray}
where $L_{f}^{(1)}$ and $L_{f}^{(2)}$ are the Vlasov
operators linearized about the one-particle distribution~$f$, and acting, respectively,
on the first pair $(q_{1},p_{1})$ and on the second pair $(q_{2},p_{2})$ of
the function $g=g(q_{1},p_{1},q_{2},p_{2},t)$.
Explicitly, for a function $h$ of $(q,p,t)$, the expression for the linear Vlasov operator $L_fh$ is
\begin{equation}
L_{f}h(q,p,t)=p\frac{\partial h}{\partial q}-f'(p,t)\int{\rm d}q_{1}{\rm d}p_{1}\,v'(q-q_{1})h(q_{1},p_{1},t),
\end{equation}
so that we have for $L_{f}^{(1)}g$,
\begin{eqnarray}\label{eq:linear-vlasov-operator}
L_{f}^{(1)}g(q_{1}-q_{2},p_{1},p_{2},t)=p_{1}\frac{\partial g}{\partial
q_{1}}-f'(p_{1},t)\int{\rm d}q_{3}{\rm
d}p_{3}\,v'(q_{1}-q_{3})g(q_{3}-q_{2},p_{3},p_{2},t). \nonumber \\
\end{eqnarray}
$L_{f}^{(2)}g$ is obtained from Eq.~(\ref{eq:linear-vlasov-operator})
by exchanging the subscripts $1$ and $2$. 

To obtain from Eqs.~(\ref{eq:stochastic-BBGKY-1-eq}) and
(\ref{eq:stochastic-BBGKY-2-eq}) a single kinetic equation for the distribution
function~$f$, we have to solve Eq.~(\ref{eq:stochastic-BBGKY-2-eq})
for $g$ as a function of $f$ and plug the result into the right
hand side of Eq.~(\ref{eq:stochastic-BBGKY-1-eq}). Because the two
equations are coupled, this program is not achievable without making further
simplifying assumption. Nevertheless, we readily see from these
equations that the two-particle correlation
$g$ evolves over a timescale of order one, whereas the one-particle
distribution function $f(p,t)$ evolves over a timescale of order
$1/\alpha$. We may then use this timescale separation and compute the
long-time limit of $g$ from Eq.~(\ref{eq:stochastic-BBGKY-2-eq})
by assuming $f$ to be steady in time; this is the equivalent of the Bogoliubov's
hypothesis in the kinetic theory for isolated systems with long-range
interactions. Note that for this timescale separation to be valid, we must also
suppose that the one-particle distribution function $f(p,t)$ is a
stable solution of the Vlasov equation at all times. Indeed, if this
is not the case, it can be shown that $g$ diverges in the limit $t \to
\infty$ \cite{Nardini-Bouchet}. The physical content of this hypothesis
is that the system slowly evolves from the initial condition through a
sequence of quasistationary states to the final stationary state.

Because we assume the system to be homogeneous in space, it is useful
to Fourier transform Eqs.~(\ref{eq:stochastic-BBGKY-1-eq}) and (\ref{eq:stochastic-BBGKY-2-eq})
with respect to the spatial variable; we get
\begin{equation}\label{eq:stochastic-BBGKY-1-eq-fourier-transform}
\frac{\partial f}{\partial t}-\alpha\frac{\partial}{\partial p}[pf]-\frac{\alpha}{2}C(0)\frac{\partial^{2}f}{\partial p^{2}}=-2\pi i\alpha\sum_{k=-\infty}^{\infty}kv_{k}\frac{\partial}{\partial p}\int\mathrm{d}p'\, g_{k}(p,p',t),
\end{equation}
and 
\begin{equation}\label{eq:stochastic-BBGKY-2-eq-fourier-transform}
\left(\frac{\partial g_{k}}{\partial
t}+L_{f,k}^{(1)}g_{k}+L_{f,-k}^{(2)}g_{k}\right)(p_{1},p_{2},t)=c_{k}f'(p_{1})f'(p_{2}),
\end{equation}
where $g_{k}(p_{1},p_{2},t)$ is the Fourier transform of $g(q,p_{1},p_{2},t)$
with respect to the spatial variable, and $v_{k}$ is the $k$-th Fourier coefficient of the pair potential
$v(q)$. The explicit expression for the
$k$-th Fourier component of the linear Vlasov operator $L_{f,k}$ acting
on a function $h(p)$ is 
\begin{equation}\label{eq:linear-vlasov-operator-fourier-transform}
\Bigl(L_{f,k}h\Bigr)(k,p)=ikph(p)-2\pi ikv_{k}f'(p)\int\mathrm{d}p'\, h(p').
\end{equation}
One has analogous expressions for $L_{f,k}^{(1)}$ and for $L_{f,-k}^{(2)}$. We readily see that $L_{f,k}^{*}=L_{f,-k}$.

From the right hand side of Eq.~(\ref{eq:stochastic-BBGKY-1-eq-fourier-transform}),
we see that to obtain a single kinetic equation,
we need only the Fourier transform $g_{k}(p,p',t)$,
more specifically, its integral with respect to the second momentum
variable $p'$. Actually, it can be shown that $g_{k}(p,p',t)$ does not
have a well-defined time-asymptotic (it converges only in
the sense of distribution), while its integral with respect to $p'$ does
have; this is connected to the mechanism of Landau damping \cite{Nardini-Bouchet}.

The structure of Eqs.~(\ref{eq:stochastic-BBGKY-1-eq}) and
(\ref{eq:stochastic-BBGKY-2-eq}), or, equivalently, of Eqs.~(\ref{eq:stochastic-BBGKY-1-eq-fourier-transform}) and
(\ref{eq:stochastic-BBGKY-2-eq-fourier-transform}) is very familiar in
kinetic theories; we refer the reader to \cite{Nardini-Bouchet} for a
general discussion. Equation~(\ref{eq:stochastic-BBGKY-2-eq}), or, equivalently,
Eq.~(\ref{eq:stochastic-BBGKY-2-eq-fourier-transform}), is called the
Lyapunov equation for the two-point correlation of a stochastic variable
described by an Ornstein--Uhlenbeck process. However, there is a difference from
the standard finite-dimensional case \cite{Gardiner} in that in our
case, $L_f$ is a linear infinite-dimensional operator acting on a
functional space, instead of being a finite-dimensional one, i.e., a
matrix. This makes it non-trivial to compute the long-time asymptotic of
the right hand side of Eq.~(\ref{eq:stochastic-BBGKY-1-eq}), where $g$
is the solution of Eq.~(\ref{eq:stochastic-BBGKY-2-eq}) with $f$
steady in time.
A possible way to achieve this goal is to follow the derivation of the
Lenard--Balescu equation from the BBGKY hierarchy, as may be found in
the Appendix A of Ref. \cite{Nicholson:1992}. For
explicit technical details, see \cite{Nardini-Bouchet}, in which the
method to solve the Lyapunov equation in a general manner is discussed,
and, subsequently, applied to the derivation of kinetic theories for
long-range interacting systems and two-dimensional turbulence models.
 
In the present case, the linear transform of the stationary solution of
the Lyapunov equation, Eq.~(\ref{eq:stochastic-BBGKY-2-eq-fourier-transform}),
which is needed to compute the right hand side of Eq.~(\ref{eq:stochastic-BBGKY-1-eq-fourier-transform}), can be written (see
Ref. \cite{Nardini-Bouchet}) in the frequency space as
\begin{eqnarray}\label{solution_lyapunov_equation}
\int dp_{1}\, g_{k}^{\infty}[f](p,p_{1})&\equiv& \lim_{t\to \infty}\int
dp_{1}\, g_{k}(p,p_{1},t) \\
&=&\frac{1}{\pi}\int_{\Gamma}\mathrm{d}\omega\,\Bigl(R_{f,k}(\omega)b\Bigr)(p)\,\int \mathrm{d}p'\,\Bigl(R_{f,-k}(-\omega)b^{*}\Bigr)(p'),
\end{eqnarray}
where $\Gamma$ is a contour which passes above all the poles of
$\Bigl(R_{f,k}(\omega)b\Bigr)$, and $R_{f,k}(\omega)$ is the resolvent operator, defined as
\begin{equation}
R_{f,k}(\omega)\equiv(-i\omega+L_{f,k})^{-1},
\end{equation}
while $b(p,t)=\sqrt{c_k}f'(p,t)$. The discussion of the explicit form of the resolvent operator is a standard topic
in plasma theory, and involves the phenomenon of Landau damping; we refer the
reader to classical references for this result, for example, 
\cite{Balescu:1997,Nicholson:1992,Lifshitz:2002}.
Its action on a function $h$, defined for $\omega$
such that $\mathrm{Im}(\omega)>0$, is 
\begin{equation}\label{Resolvent_operator}
\Bigl(R_{f,k}(\omega)h\Bigr)(p)=\frac{1}{-i\omega+ikp}\left[h(p)+\frac{2\pi
ikv_{k}}{\epsilon(k,\omega)}f'(p)\int\mathrm{d}p'\,\frac{h(p')}{-i\omega+ikp'}\right],
\end{equation}
where $\epsilon(k,\omega)$ is the dielectric function, which for $\mathrm{Im}(\omega)>0$ is given by 
\begin{equation}\label{eq:stochastic-dielectric-function}
\epsilon(k,\omega)=\left[1-2\pi iv_{k}k\int{\rm d}p\frac{f'(p)}{-i\omega+ikp}\right],
\end{equation}
and by its analytic continuation for $\omega$ when $\mathrm{Im}(\omega)\leq 0$. 
Both the resolvent operator and the dielectric function are defined for
$\omega\in\mathbb{R}$ by their analytic continuation, which will still be denoted by the same symbols.

Now, inserting Eq.~(\ref{Resolvent_operator}) into Eq.~(\ref{solution_lyapunov_equation}), and with some calculations whose details will be reported in \cite{Nardini-Bouchet}, we get the kinetic equation
\begin{equation}\label{eq:stochastic-kinetic-equation}
\frac{\partial f}{\partial t}-\alpha\frac{\partial(pf)}{\partial
p}-\alpha\frac{\partial}{\partial p}\left[D[f]\frac{\partial f}{\partial
p}\right]=0,
\label{kinetic_equation}
\end{equation}
where 
\begin{eqnarray}\label{diffusion_coefficient}
D[f](p)&=&\frac{1}{2}C(0)\nonumber \\
&&+2\pi\sum_{k=1}^{\infty}v_{k}c_{k}\int^{*}{\rm d}p_{1}\left[\frac{1}{|\epsilon(k,kp)|^{2}}+\frac{1}{|\epsilon(k,kp_{1})|^{2}}\right]\frac{1}{p_{1}-p}f'(p_{1},t).
\end{eqnarray}
We recall that $v_{k}$ is the $k$-th Fourier coefficient of the pair potential
$v(q)$, the quantity $c_{k}$ is defined in Eq.~(\ref{ckdefinition}),
$\epsilon$ is the dielectric function defined in Eq.~(\ref{eq:stochastic-dielectric-function}), and $\int^{*}$ indicates the Cauchy integral or Principal Value.

The kinetic equation (\ref{eq:stochastic-kinetic-equation}) is the
central result of the kinetic theory developed in this paper. It has the
form of a non-linear Fokker-Planck equation \cite{Risken} because the
diffusion coefficient $D[f](p)$ is itself a functional of the
one-particle distribution function $f$. The linear part of the diffusion
coefficient $(1/2)C(0)$ is the mean-field effect of the stochastic
forces, whereas the effect of two-particle correlation is encoded in
the non-linear part. In the next section, we describe how we use this
kinetic equation to get information about the nonequilibrium stationary
states of the dynamics.

In the foregoing, we discussed the kinetic theory in the limit $N\alpha\gg1$.
The extension to the general case is straightforward: Because of the
linearity of the equations of the hierarchy
(\ref{eq:stochastic-BBGKY-1-eq}) and (\ref{eq:stochastic-BBGKY-2-eq}),
the finite-$N$ and stochastic effects give independent contributions.
The kinetic equation at leading order of both stochastic and finite-size
effects is 
\begin{equation}
\frac{\partial f}{\partial t}=Q_{\alpha}[f]+Q_{N}[f],
\end{equation}
where $Q_{\alpha}$ is the operator described in Eq.~(\ref{eq:stochastic-kinetic-equation}),
and $Q_{N}$ (of order $1/N$) is the Lenard-Balescu operator. For
instance, in the case $N\alpha\ll1$ and in dimensions greater than
one, the operator~$Q_{N}$ is responsible for the relaxation to Boltzmann
equilibrium after a timescale of order $N$, whereas the smaller effect
of $Q_{\alpha}$ selects the actual temperature after a longer timescale
of order $1/\alpha$. 
\subsection{Numerical simulations}
\label{numerical-scheme}
\begin{figure}[here]
\begin{center}
\includegraphics[width=90mm]{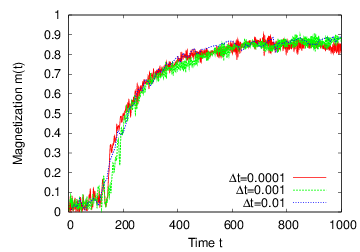}
\end{center}
\caption{Magnetization as a function of time, obtained from numerical
simulation of the stochastically-forced HMF model with
$N=1000,\alpha=0.01$ at kinetic temperature $T=0.25$, and with
$c_1=c_2=c_3=\ldots=c_{10}=C(0)/20, c_{k \ge
11}=0$, where $C(0)=2T$. The
values of the integration step size $\Delta t$ used 
are marked in the figure. The data are
obtained by using the integration algorithm described in section \ref{numerical-scheme}. That the magnetization plots collapse onto one curve shows the
stability of our algorithm with respect to variation
in $\Delta t$. We have checked that the final value of the magnetization
matches with the prediction from equilibrium statistical mechanics.}
\label{algo-check-fig}
\end{figure}
Here we describe how we may simulate the dynamics
(\ref{equations_of_motion}) by means of a numerical integration scheme.
To simulate the dynamics over a given time interval $[0:\mathcal{T}]$, choose
a time step size $\Delta t$, and set $t_n=n\Delta t$ as the $n$-th time step of the
dynamics. Here, $n=0,1,2,\ldots,N_t$, where $N_t=\mathcal{T}/\Delta t$. In our
numerical scheme, at every time step, we first
discard the effect of the noise and employ a fourth-order symplectic
algorithm to integrate the deterministic Hamiltonian part of the
dynamics \cite{McLachlan}.
Subsequently, we add the effect of noise and implement an Euler-like
first-order algorithm to update the dynamical variables \footnote{We
found that an Euler-like first-order scheme alone is unstable with respect to not-too-small 
$\Delta t$, in the sense that one obtains different magnetization profiles as
a function of time $t=t_n \Delta t$. The situation gets worse for small $\alpha$, when
one needs to use very small $\Delta t$ to obtain consistent results.
Therefore, for faster and efficient simulation, we adopted the ``mixed"
scheme described in the text.}.   
Specifically, one step of
the scheme from $t_n$ to $t_{n+1}=t_n+\Delta t$ involves the following
updates of the dynamical variables for $i=1,2,\ldots,N$: For the
symplectic part, we have, for $m=1,\ldots,4$, 
\begin{eqnarray}
&&p_i\Big(t_{n}+\frac{m\Delta t}{4}\Big)=p_i\Big(t_n+\frac{(m-1)\Delta
t}{4}\Big)+b(m)\Delta t
\Big[-\frac{\partial H}{\partial q_i}(\{q_i\Big(t_n+\frac{(m-1)\Delta
t}{4}\Big)\})\Big], \nonumber \\
\label{formalintegration3} \\
&&q_i\Big(t_{n}+\frac{m\Delta t}{4}\Big)=q_i\Big(t_n+\frac{(m-1)\Delta
t}{4}\Big)+a(m)\Delta t ~p_i\Big(t_n+\frac{m\Delta t}{4}\Big),\nonumber 
\end{eqnarray}
where the constants $a(m)$'s and $b(m)$'s are given in Ref.
\cite{McLachlan}.
At the end of the update (\ref{formalintegration3}), we have the set
$\{q_i(t_{n+1}),p_i(t_{n+1})\}$.
Next, one includes the effect of the
stochastic noise by leaving $q_i(t_{n+1})$'s unchanged,  but by
updating $p_i(t_{n+1})$'s as
\begin{eqnarray}
&&p_i(t_{n+1}) \to p_i(t_{n+1})\Big[1-\alpha \Delta t\Big]+\sqrt{\alpha}\Big[\sqrt{c_0}\Delta
X^{(0)}(t_{n+1})\nonumber \\
&&+\sum_{k=1}^{N_R} \sqrt{2c_k} \Big\{\Delta
X^{(k)}(t_{n+1})\cos\Big(kq_i(t_{n+1})\Big)+\Delta
Y^{(k)}(t_{n+1})\sin\Big(kq_i(t_{n+1})\Big)\Big\}\Big].
\label{formalintegration4}
\end{eqnarray}
Here $\Delta X^{(k)}$ and $\Delta Y^{(k)}$ are Gaussian distributed
random numbers with zero mean and unit variance.
The outcome of implementing this mixed scheme for the
stochastically-forced HMF model is shown in Fig.
\ref{algo-check-fig}, where one may observe consistent
results with respect to change of $\Delta t$ over a wide range of values. 
In numerical simulations reported later in the paper, we exclusively
used
this mixed scheme to simulate the dynamics (\ref{equations_of_motion}).

%=============================================================================
\section{Predictions of the kinetic theory and comparison with
simulations}
\label{results1}
\begin{figure}[here]
\begin{center}
\includegraphics[width=80mm]{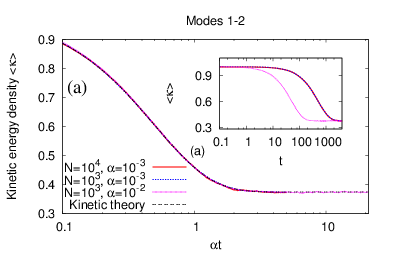}
\includegraphics[width=80mm]{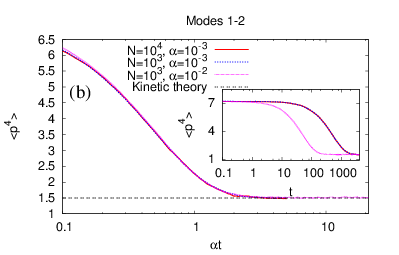}
\end{center}
\caption{(a) Kinetic energy density $\langle \kappa
\rangle$, and (b) $\langle p^4\rangle$ as a
function of $\alpha t$, at kinetic temperature $T=0.75$, with modes
$1-2$ excited with amplitudes satisfying $c_1=c_2=C(0)/4$, where $C(0)=2T$. The data for different $N$ and $\alpha$
values are obtained from numerical simulations of the
stochastically-forced HMF model with $\Delta t=0.01$, and involve averaging over $50$ histories for $N=10^4$
and $10^3$ histories
for $N=10^3$. The data collapse implies that $\alpha$ is the timescale
of relaxation to the stationary state. The inset shows the data without
time rescaling by $\alpha$. Similar plots for different
parameter values were reported in Ref. \cite{Nardini:2012}.}
\label{p2p4}
\end{figure}

\begin{figure}[here]
\begin{center}
\includegraphics[width=80mm]{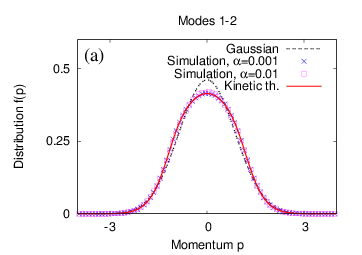}
\includegraphics[width=80mm]{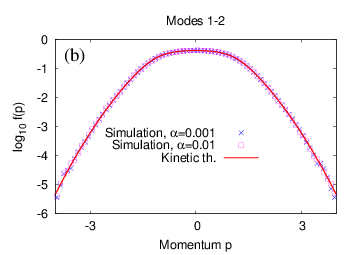}
\end{center}
\caption{Stationary momentum distribution $f(p)$, on (a) linear, and (b) semi-log scales, for $\alpha=0.001$, and $0.01$ at kinetic
temperature $T=0.75$. The plots correspond to modes $1$ and $2$ excited
with amplitudes satisfying $c_1=c_2=C(0)/4$, where $C(0)=2T$. The data denoted by crosses and
squares are results of $N$-body simulations of the stochastically-forced HMF model
with $N=10000, \Delta t=0.01$ and $1000$ independent realizations of the
dynamics, while the red continuous lines refer to the theoretical
prediction from the kinetic theory. For comparison, the black broken line shows the Gaussian
distribution with the same kinetic energy (stationary state of the
stochastically-forced HMF model at $T=0.75$, $c_0=c_1=0,c_2=0.75,c_{k
\ge 3}=0$).}
\label{fp}
\end{figure}

\begin{figure}[here]
\begin{center}
\includegraphics[width=80mm]{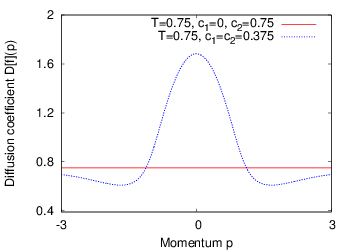}\\
\end{center}
\caption{Diffusion
coefficient $D[f](p)$ for the stationary momentum distribution
$f(p)$ at kinetic temperature $T=0.75$, with $c_0=c_{k \ge 3}=0$, and either
(i) $c_1=c_2=0.375$, or, (ii) $c_1=0, c_2=0.75$.}
\label{Dp}
\end{figure}

We now focus on how to obtain from the kinetic equation
(\ref{eq:stochastic-kinetic-equation}) predictions for the
nonequilibrium stationary states of the system. According to Eq.~(\ref{eq:stochastic-kinetic-equation}), $1/\alpha$ is only a timescale;
thus, at leading order in $\alpha$ and except for a time rescaling, the
parameter $\alpha$ 
does not affect the time evolution of the system. This statement holds also beyond
the leading order for what concerns the evolution of the kinetic energy;
its evolution may be obtained directly from the equations of motion
(\ref{equations_of_motion}), as discussed in section \ref{model}, and
can also be obtained from the kinetic equation
(\ref{eq:stochastic-kinetic-equation}), as detailed in \ref{kinetic-energy-evolution}. For the evolution of
other observables, there will be corrections at higher orders in $\alpha$. 

As previously discussed, the dynamics of the system does not respect 
detailed balance if the forcing is not white in space. At the
level of the kinetic equation, by inspecting the definition of the
diffusion coefficient, Eq.~(\ref{diffusion_coefficient}), we see that
the effect of correlations induced by the stochastic forces is modulated
by the Fourier component $v_k$ of the interparticle potential. Then,
taking the forcing spectral amplitudes $c_k$ different from zero if and
only if $v_k=0$, the non-linear part of the diffusion coefficient
vanishes. On the other hand, taking $c_k \ne 0$ for the modes for which
$v_k \ne 0$ leads to a diffusion constant which has a non-vanishing
non-linear part. To be concrete, let us discuss these two scenarios in the context of the stochastically-forced HMF
model.

Since the Fourier transform of the HMF interparticle potential is, for $k \neq 0$,
$v_k=-\left[\delta_{k,1}+\delta_{k,-1}\right]/2$, it follows that only the stochastic force mode with wave number
$k=1$ contributes to the non-linear part of the diffusion coefficient;
all the other stochastic force modes result in only a
mean-field contribution through the term
$C(0)$. Thus, for the case $c_1 \ne 0$, the two relevant parameters that dictate
the evolution of the
stochastically-forced HMF model by the kinetic equation
(\ref{eq:stochastic-kinetic-equation}) with a non-linear diffusion
coefficient are $C(0)$ and~$c_1$. From Eq.~(\ref{kinetic-temp-defn}), since $C(0)$ is related to the kinetic
temperature $T$, we take $T$ and $c_1$ to be the two relevant
parameters.
From Eq.~(\ref{kinetic-energy-steady-state}), we know that $2T$ equals
the kinetic energy in the final stationary state.
Also, Eq.~(\ref{eq:g}) implies that $c_1 \le C(0)/2$. 

If however $c_1=0$, then, at leading order
in $\alpha$, the dynamics of the system is described by a linear
Fokker-Planck equation; this equation is the same as the one which
describes the HMF system when coupled to a Langevin thermostat,
studied in~\cite{ChavanisBMF,Baldovin_Orlandini}. This means
that for this particular choice of the parameters, the detailed balance
is broken for the dynamics, but this feature cannot be seen in the
kinetic theory, being an effect at a higher order in~$\alpha$. 
In this case, the homogeneous stationary states of the kinetic
equation have Gaussian momentum distribution $f(p)$.  As has been studied
thoroughly in the context of canonical equilibrium of the HMF model, these states
are stable for kinetic energies greater than $1/4$, i.e., for $C(0)>1$. 

Except for the special case of $c_1 =0$, the stationary velocity
distribution of the kinetic equation
(\ref{eq:stochastic-kinetic-equation}) is in general not Gaussian. This
can be seen semi-analytically by observing that the Gaussian distribution function
\begin{equation}\label{Gaussian}
f_G(p)=A\,\exp(-\beta p^2)\,,\qquad A=\sqrt{\frac{\beta}{\pi}}\,,\qquad
\beta=\frac{1}{2T},
\end{equation}
with $\beta$ chosen such that the value of the kinetic energy is the one
selected by $T$, solves the linear Fokker-Planck equation with the diffusion coefficient given by
\begin{equation}
D_{mf}=T.
\end{equation}
To prove that the Gaussian distribution function is not a stationary
solution of Eq.~(\ref{eq:stochastic-kinetic-equation}), we have to prove that the contribution to $\partial f/\partial
t$ from the non-linear part of $D[f](p)$ in Eq.~(\ref{diffusion_coefficient}) does not vanish. This result can be proven
with an asymptotic expansion \cite{Lifshitz:2002} for large momenta of the integrals which
appear in the diffusion coefficient. We report the
straightforward computation in \ref{nonGaussian}. From the same
analysis, one can deduce that, even though the distribution function is not Gaussian, its tails are Gaussian.

On the basis of the above discussions, we expect that for values of $T$
and $c_1$ such that $T>0.5$ and $c_1\ll 2T$,
the stationary states will be close to homogeneous states
with Gaussian momentum. In order to locate the actual stationary states of the kinetic
equation, we have devised a simple numerical scheme, based on the observation that a linear Fokker-Planck equation whose diffusion coefficient $D(p)$ is strictly positive admits a unique stationary state 
\begin{equation}
\label{stationary-FokkerPlanck}
f_{ss}(p)=A\exp\left[-\int_0^p {\rm d}p' ~\frac{p'}{D(p')}\right].
\end{equation}
For a given distribution $f_n(p)$, we compute the diffusion coefficient
$D_n(p)$ through Eq.~(\ref{diffusion_coefficient}), and then $f_{n+1}$ using $D_n$ and
Eq.~(\ref{stationary-FokkerPlanck}). This procedure defines an iterative
scheme. Whenever convergent, this scheme leads to a stationary state of
Eq.~(\ref{eq:stochastic-kinetic-equation}). Each iteration involves
integrations, so that we expect the method to be robust enough when starting not too far from an actual stationary state. However, we have no detailed mathematical analysis yet. 
Implementing this iterative scheme, we observed that the distribution
$f_{\infty}$ to which the scheme converges is independent of the initial
distribution $f_0$. Moreover, the convergence time is exponential in the
number of steps $n$ whenever $T$ is not too close to loss of stability
of $f_{\infty}$ with respect to the linear Vlasov dynamics; in practice,
we are able to get reliable results for $T \gtrsim 0.65$. 

In order to check the theoretical predictions discussed above, we
performed numerical simulations of the stochastically forced HMF model.
Figure \ref{p2p4} shows the evolution of the kinetic energy and $\langle p^4
\rangle=(1/N)\sum_{i=1}^N p_i ^4$, where they have been compared with
our theoretical predictions. In the case of $\langle p^4 \rangle$, we have
compared the long-time asymptotic value with the kinetic theory
prediction for the stationary state, computed numerically by using the
iterative solution for the stationary distribution. The figure
illustrates a very good agreement between the theory and simulations.
For a more accurate check of the agreement, we have obtained the stationary momentum distribution from both $N$-body simulations and the numerical
iterative scheme discussed above. A comparison between the two, shown in
Fig. \ref{fp}, both on linear and semi-log scales, shows a very good
agreement between theory and simulations. In this figure, we also show
the Gaussian distribution with the same kinetic energy, to illustrate
the point that the stationary momentum distribution of the system is far
from being Gaussian. The diffusion coefficient $D[f](p)$ is shown in 
figure \ref{Dp}.

In passing, let us remark that, with an iterative scheme analogous to the one
described above, one could have also obtained the full time evolution $f(p,t)$
that obeys the kinetic equation (\ref{eq:stochastic-kinetic-equation}). However, we will not address this point here.

We also note that while a linear Fokker-Planck equation with non-degenerate diffusion
coefficient can be proven to converge to a unique stationary
distribution~\cite{Risken}, this is not true in general for non-linear
Fokker-Planck equations like Eq.~(\ref{eq:stochastic-kinetic-equation}). We expect that if the dynamics is not too far from detailed balance, the kinetic equation will have
a unique stationary state.  Far from equilibrium, the kinetic equation could lead to very
interesting dynamical phenomena, like bistability, limit cycle or more
complex behaviors. The main issue is then the analysis of the evolution
of the kinetic equation.  
Although some methods to
study this type of equation exist \cite{Frank}, we have only the numerical iterative scheme described above to provide some preliminary answers. A more rigorous
mathematical analysis is left for future studies.

\section{Nonequilibrium phase transition and collapse}
\label{results2}
Until now, we have considered homogeneous stationary states of the
dynamics (\ref{dynamics}), and have discussed a kinetic theory to analyze
them. Although our theory can in principle be extended to include
inhomogeneous stationary states, its actual implementation to get, e.g.,
the single-particle distribution, would require more involved
computations than the one we encountered for homogeneous states. In order to get preliminary answers, we have
performed extensive numerical simulations of the dynamics in the context
of the stochastically-forced HMF model. Our specific interest is  
to know about how the magnetization behaves as the kinetic temperature is
reduced from high values. 
\begin{figure}[here]
\begin{center}
\includegraphics[width=80mm]{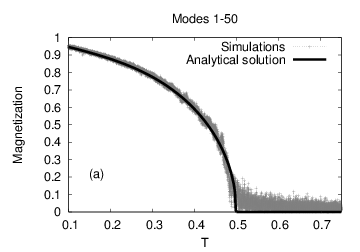}
\includegraphics[width=80mm]{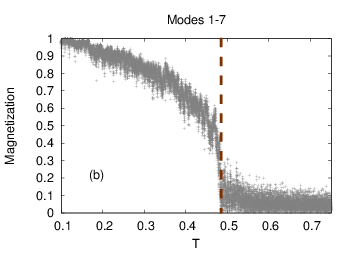}
\includegraphics[width=80mm]{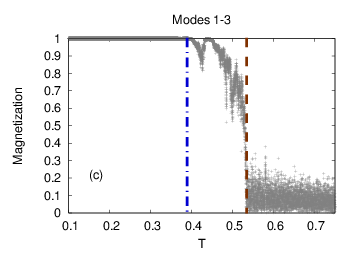}
\includegraphics[width=80mm]{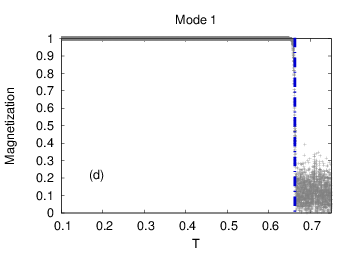}
\end{center}
\caption{Numerical simulation results for magnetization in the stochastically-forced
HMF model as a function of adiabatically-tuned kinetic temperature $T$; the different plots correspond
to different number of modes excited in the spectrum with
amplitudes satisfying $C(0)=c_0+2\sum_{k=1}^{N_R}c_k$, where $C(0)=2T$,
the index $k=1,2,\ldots,N_R$ denotes the mode number, while $N_R$ is the total number of excited modes with $k \ne 0$. In all cases,
the modes excited were chosen to have equal amplitudes, with $c_0=0, N=5000, \alpha=0.01, \Delta t=0.01$, while the tuning rate for $T$
is $10^{-5}$. It may be noted that forcing
equally a large number of modes ($\sim 50$) reproduces the equilibrium
magnetization profile as illustrated by the match with the analytical
equilibrium solution in the panel (a). In panel (b), the first-order
nonequilibrium phase transition is marked by the vertical dashed line.
In panel (c), besides the first-order transition, we also show the
dynamical transition to the collapsed state by the vertical dashed dotted line. In panel
(d), the nonequilibrium phase transition and the dynamical transition almost
coincide, and we just show the latter one by the vertical dashed dotted
line.}
\label{inhom-1}
\end{figure}

In the case when the stochastic forcing
respects detailed balance (i.e., when the noise spectrum is flat and all modes 
are excited), the stochastically-forced HMF model reduces to the
Brownian mean-field (BMF) model studied previously \cite{ChavanisBMF}. Here, we know that the system settles into an equilibrium state in which
it exhibits a second-order phase transition
at the kinetic temperature $T=T_c=1/2$: on increasing $T$ from low
values, the magnetization decreases
continuously to zero at $T_c$ and remains zero at higher temperatures.
In the following, we excite only a limited number of modes $N_R$, but
the amplitudes of all excited modes are equal ($c_k$ equals $c$ for all
$k \leq N_R$, and is zero otherwise, where the constant $c$ is related to the temperature). Figure \ref{inhom-1}(a) shows that with $N_R=50$, one reproduces very well
the equilibrium profile of the magnetization as a function of
temperature. On reducing the value of $N_R$, the system is
driven more and more out of equilibrium. Indeed, Fig. \ref{inhom-1}(b) shows
that with $N_R=7$, the magnetization profile changes; in particular, it
develops a discontinuity around a temperature $T_{trans} \approx 0.49$,
reminiscent of a first-order phase transition.
The transition temperature is denoted by the vertical dashed line. With $N_R=3$, Fig. \ref{inhom-1}(c) shows that the discontinuity gets more
pronounced, and $T_{trans}$ is now shifted to a higher
value (denoted again by the vertical dashed line). A new feature appears
in this plot, namely, at a temperature $T_{dyn} \approx 0.4$, 
the magnetization attains the maximal value of unity, which it retains for
all lower temperatures. This value of unity corresponds to a state in
which the particles are very close to one another on the circle, thus
defining a ``collapsed" state.  We found that this state, as well
as the transition to it, persist on changing the system size $N$. 

Now, it is known that trajectories of ensembles of dissipative dynamical
systems forced by the same realization of a stochastic noise converge to
a single one \cite{Arnold,Maritan-etal}. These attracting trajectories
are referred to as the ones due to the so-called stochastic attractor. 
Although we did not perform a detailed characterization of the collapse
in our model, we believe that the phenomenon is related to stochastic attractors.  

Coming back to Fig.
\ref{inhom-1}(c), we see that for temperatures $T_{dyn} < T <T_{trans}$, the
magnetization shows strong fluctuations. Reducing the number of excited
modes to a single one, namely, to the one that coincides with the Fourier
mode of the HMF potential, it seems from Fig. \ref{inhom-1}(d) that only
the dynamical transition to the collapsed state at a temperature
$T_{dyn} \approx 0.66$ persists. 

\begin{figure}[here]
\begin{center}
\includegraphics[width=80mm]{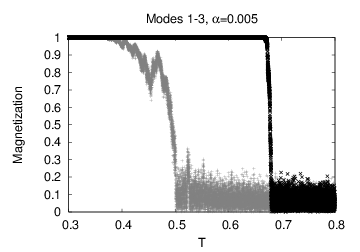}
\includegraphics[width=80mm]{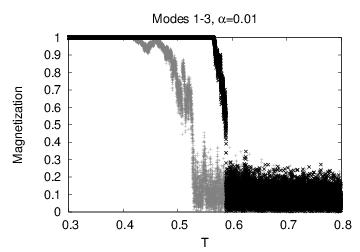}
\end{center}
\caption{Numerical simulation results for magnetization in the stochastically-forced HMF
model as a function of adiabatically-tuned kinetic temperature $T$ for
two different values of $\alpha$. In each case, the modes $1-3$ are
excited with amplitudes satisfying $c_1 = c_2 = c_3 = C(0)/6$, where
$C(0) = 2T$. In all cases, $N = 5000, \Delta t = 0.01$, while the tuning rate
for $T$ is $10^{-5}$. The grey points correspond to the case when the temperature is decreased from high values, while
the black points correspond to the case when the temperature is
increased from low values.}
\label{hys}
\end{figure}

The hint that the nature of the phase transition at $T_{trans}$ is of
first-order comes from the hysteresis plots of 
Fig. \ref{hys}. To obtain these plots, one monitors the magnetization
while tuning adiabatically the kinetic temperature across $T_{trans}$
from higher to lower values and back to complete a full cycle.
As is evident from Fig. \ref{hys}, the observed hysteresis is between the collapsed state and the zero-magnetization state. In principle, it should be possible 
to observe a hysteretical behavior between the magnetized and the zero-magnetization state. To achieve this in simulations involving adiabatic tuning of temperature, one should not allow the system to make the transition to the collapsed state, which requires conditions close to those that ensure detailed balance. However, a possible drawback of this method is that closeness to detailed balance might lead to narrow hysteresis loops. Moreover, the adiabatic tuning of temperature should not be very slow, as otherwise one observes bistability instead of the hysteresis. All these factors make the observation of hysteresis between the magnetized and the zero-magnetization state difficult to observe numerically; further explorations of this will be the subject of future investigations.

\begin{figure}[here]
\begin{center}
\includegraphics[width=8cm]{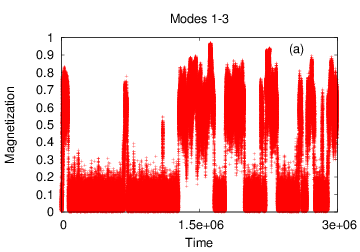}
\includegraphics[width=80mm]{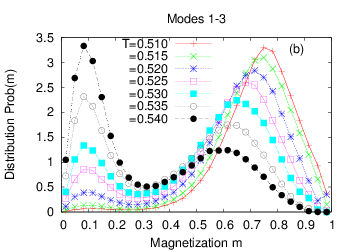}
\end{center}
\caption{(a) Numerical simulation results for magnetization in the stochastically-forced
HMF model as a function of time at kinetic temperature
$T=0.53$, with $N=5000, \alpha=0.005,
\Delta t=0.01$, and with modes $1-3$ excited, whose amplitudes satisfy
$c_1=c_2=c_3=C(0)/6$, where $C(0)=2T$. The figure shows clear signatures of bistability
in which the system during the course of evolution switches back and
forth between spatially homogeneous ($m \sim O(0)$) and inhomogeneous
($m \sim O(1)$) states. (b) Distribution $Prob(m)$ of the magnetization $m$ as a function of
$T$ at a fixed value of $\alpha=0.01$. The data are obtained from
numerical simulation results similar to (a) for magnetization in the stochastically-forced
HMF model, with $N=5000, \Delta t=0.01$, and with modes $1-3$ excited, whose amplitudes satisfy
$c_1=c_2=c_3=C(0)/6$, where $C(0)=2T$.}
\label{inhom-2}
\end{figure}

In order to explore further the region in Fig. \ref{inhom-1}(c) close to
$T_{trans}$, and to ascertain the nature of the phase transition
at $T_{trans}$, we fix the value of the temperature to be $T=0.53$, and monitor the
magnetization as a function of time. The time series of the
magnetization is shown in Fig. \ref{inhom-2}(a), in which one observes
clear signatures of bistability, whereby the system switches back and
forth between homogeneous ($m \approx 0$) and inhomogeneous ($m >0$) states. In addition, we show
in Fig. \ref{inhom-2}(b) the distribution of the magnetization around
the phase transition temperature: the distribution is bimodal with a peak
 around a zero value and another one around a positive value. When
 decreasing the temperature across the phase transition region, we
 clearly see that the peak heights of the distributions of the
 magnetization at the zero and
 non-zero values interchange. These two features, together with the
 hysteresis plots of Fig. \ref{hys}, support the first-order
 nature of the transition around $T_{trans}$ which can be estimated from
 Fig. \ref{inhom-2}(b) to be $T_{trans} \approx 0.532$.

\begin{figure}[here!]
\centering
\includegraphics[width=80mm]{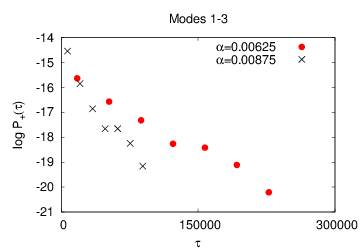}
\caption{Distribution of the residence time $\tau$ in the inhomogeneous
state, for two
values of $\alpha$. The data are obtained from simulations with modes $1-3$
excited, whose amplitudes satisfy
$c_1=c_2=c_3=C(0)/6$, where $C(0)=2T$. Here, the kinetic temperature
$T=0.53$, while $\Delta t=10^{-2},N=5000$.}
\l{Ptau}
\end{figure}

From Fig.  \ref{inhom-2}, it is clear that the system has two well
separated attractors, corresponding to the homogeneous and inhomogeneous
states.  A question of immediate interest is: How long does the system
stay in one state before switching to the other? Let us define the 
residence time as the time the system stays in one state before it
switches to the other. In the limit of low noise level $\alpha$, there
is a clear separation between the natural dynamical time and the typical
residence time, as is evident from Fig.  \ref{inhom-2}(a). As a result, one may
conjecture that two successive switching events are statistically
independent of one another. In case such a conjecture holds for our
model, the residence time statistics will be a Poisson process,
characterized solely by the probability per unit time, $\lambda_+$, of
switching from the inhomogeneous state to the homogeneous state, and the
probability per unit time, $\lambda_-$, for the reverse switch. The distribution of residence time $\tau$ in each phase is then exponential: 
\be
P_\pm(\tau)=\fr{1}{\lambda_\pm}\exp(-\lambda_\pm \tau),
\ee
so that the average residence times in the two states are $\tau^{res}_\pm =
\fr{1}{\lambda_\pm}$. 
Such an exponential form of the residence time distribution is verified
from our simulation data displayed in Fig. \ref{Ptau}. Note that generating
such a plot requires running simulations of the dynamics for long enough
times so that the magnetization switches back and forth between the two
states a sufficient number of times, and one has good statistics
for the residence times. For low values of $\alpha$, such as those used in Fig.
\ref{Ptau}, this was often not feasible due to very long simulation
times. This results in bad statistics, and hence, the form of the plot
displayed in Fig. \ref{Ptau}, which, though good, may be improved upon
by running longer simulations. We conclude that our conjecture of two
successive jumps being independent holds for our model, and that the
average residence time fully characterizes the switching process for small enough $\alpha$. 

We now discuss how the residence times depend on the system parameters, in
particular, on $\alpha$. For an equilibrium system, the type of
switching process described above is an activation process with a
residence time described by the Arrhenius law. A simple model of such an
activation process is the Langevin dynamics of a Hamiltonian system in a
potential $\mathcal{V}$. The noise level is then related to the
temperature, and the Arrhenius law takes the form
\cite{Kramers,Gardiner,vanKampen,Hanggi} 
\begin{eqnarray}
\tau^{res}_+ \propto \exp (\Delta \mathcal{V}_{+-}/ \alpha), 
\label{Tres-expect1} \\
\tau^{res}_- \propto \exp (\Delta \mathcal{V}_{-+}/ \alpha).
\label{Tres-expect2}
\end{eqnarray}
Here, $\Delta \mathcal{V}_{+-}$ and $\Delta \mathcal{V}_{-+}$ are
respectively the potential energy barrier as observed from the
inhomogeneous and the homogeneous state.  
In a non-equilibrium context such as ours, there is no obvious equivalent of a
potential, but the law given by Eqs. (\ref{Tres-expect1}) and
(\ref{Tres-expect2}) is expected to hold on a
fairly general basis, in the limit of small noise. This may be
established from the instanton theory, or, from the Freidlin-Wentzell theory, which allows to compute
$\mathcal{V}$ explicitly for a given model \cite{Freidlin,Hugo}. Our system does not fulfill
the hypothesis of Freidlin-Wentzell theory, nevertheless, it is
interesting to check if the law given by Eqs. (\ref{Tres-expect1}) and
(\ref{Tres-expect2}) holds.
Our simulation data shown in Fig. \ref{mean-tau} show that 
the dependence of $\tau^{res}_\pm$ on $\alpha$, as in Eqs.
(\ref{Tres-expect1}) and (\ref{Tres-expect2}), holds also for our model, thereby suggesting that in the limit of low noise, the system
behaves as one with transitions activated by a weak noise.  

\begin{figure}[here!]
\centering
\includegraphics[width=80mm]{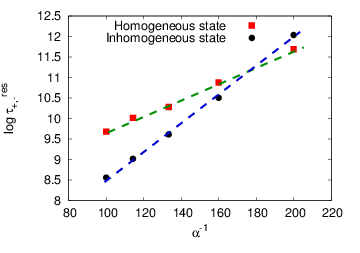}
\caption{The plot shows as a function of $1/\alpha$ the $\log$ of the mean residence
time $\tau^{res}_{+,-}$ in the two bistable states, namely, the inhomogeneous ($m
>0$) state and the homogeneous state ($m \approx 0$). The plot is based on data obtained from simulations with modes $1-3$
excited, whose amplitudes satisfy $c_1=c_2=c_3=C(0)/6$, where $C(0)=2T$. Here, the kinetic temperature
$T=0.53$, while $\Delta t=10^{-2},N=5000$. The straight line fits imply
that $\tau^{res}_{+,-} \sim  \exp(1/\alpha)$, in accordance with
Eqs.~(\ref{Tres-expect1}) and (\ref{Tres-expect2}). That the slopes of the two straight lines in
the plot are different could be due to the fact that the height of the
barrier is different when observed from the inhomogeneous
and the homogeneous state.}
\l{mean-tau}
\end{figure}

We conclude this section by describing briefly the algorithm to find
$P_{\pm}(\tau^\pm)$ to produce Fig. \ref{Ptau}, and $\tau^{res}_\pm$
to produce Fig. \ref{mean-tau}. To this end, one has to
identify from the time series data of the magnetization (see Fig.
\ref{inhom-2}(a) for an example) the switching time instants between
the two states. In the limit of very small $\alpha$, the distinction between
two states should be obvious. However, we could not reach such a limit
in our numerical simulations because the simulation time grows
exponentially with $1/\alpha$ (see Fig.~\ref{mean-tau}). For intermediate values
of $\alpha$, it is then a challenge to define precisely the two states.  Indeed, as may be seen in Fig. \ref{inhom-2}(a), the data show strong fluctuations and hence, one needs to filter out ``spurious" switching events and retain only the genuine ones.
This may be done efficiently as we now discuss. 

We first obtain from
the data a rough estimate of the mean of the magnetization when the
system is in the two states, the homogeneous and the inhomogeneous one. Let us denote by $m_>$  and $m_<$ these
estimates when the system is in the inhomogeneous and the homogeneous state,
respectively. Let us define a ``threshold" value of the magnetization
$m_{th}$ as the average of $m_>$  and $m_<$; magnetization crossing this
threshold to switch from one state to another however is not a precise enough
criterion to define a switching event, as is obvious from Fig.
\ref{inhom-2}. We thus resort to our algorithm which we now illustrate for the case when the
system is in the homogeneous state; when the system is in the
inhomogeneous state, the algorithm may be defined in a manner similar to
the one below. In our algorithm, we identify a switching event as the one for
which the following two conditions are satisfied, namely, (i) that the
magnetization crosses the threshold to switch from the homogeneous to
the inhomogeneous state, (ii) the magnetization after the switching 
reaches the value $m_>$ before reaching the value $m_<$. When a
switching event occurs, the switching time is defined as the time at
which the magnetization crossed the threshold. This algorithm allows us
to precisely define the switching times, from which we compute the
switching time statistics $P_\pm(\tau_{\pm})$, and hence, the mean residence
time $\tau^{res}_\pm$.  

%=============================================================================
\section{Conclusions}
In this work, we considered long-range interacting systems driven by
external stochastic fields, thereby leading to generic nonequilibrium
stationary states. To study spatially homogeneous stationary states, we
developed a kinetic theory approach by generalizing the known results
for isolated long-range systems. Our theoretical approach is quite general,
being applicable to any long-range inter-particle potential, space dimensions and boundary conditions.
Our extensive numerical simulations on a paradigmatic model of 
long-range interacting systems demonstrated a very good agreement with
the theory. Besides, our simulations for
this representative case illustrated very interesting bistable behavior between
homogeneous and inhomogeneous states, with a mean residence time that
diverges as an exponential in the inverse of the strength of the external stochastic
forces in the limit of low values of such forces.  

Let us note that another route to deriving the kinetic theory studied in this paper is to adopt an approach similar to the
one due to Klimontovich for isolated
systems, by writing down the time evolution equation for the noise-averaged empirical measure
$\rho(p,q,t)=(1/N)\sum_{i=1}^N \langle \delta(q_i(t)-q)\delta(p_i(t)-p)
\rangle$. In the resulting equation, the noise appears as a
multiplicative term, which can be treated perturbatively, leading to the
kinetic equation~(\ref{kinetic_equation}).  

This work leaves open some interesting issues, e.g., for technical simplicity, we assumed a
homogeneous stationary state for the development of the kinetic theory.
It would be of interest to generalize the theory to inhomogeneous
states; in this regard, the method due to Heyvaerts
reported recently may come of help \cite{Heyvaerts}. Another issue is to study the dynamics of the kinetic
equation (\ref{kinetic_equation}), both analytically and numerically,
which may unveil very interesting behaviors, such as limit cycles. One
may also hope to develop a kinetic theory similar to the one
analyzed here for related
systems, for example, the point vortex model and the Euler equations
in two-dimensional turbulence \cite{Bouchet-Venaille}.  

%=============================================================================
\section{Acknowledgments}
C. N. acknowledges the EGIDE scholarship funded by Minist\`ere
des Affaires \'Etrang\`eres. S. G. and S. R. acknowledge
the contract LORIS (ANR-10-CEXC-010-01). F. B. acknowledges the ANR
program STATOCEAN (ANR-09-SYSC-014). Numerical simulations were done at
PSMN, ENS-Lyon. S. G. is grateful to the Korea Institute for Advanced
Study (KIAS), Seoul for the kind hospitality during the time part of this work
was being done and written up.
%=============================================================================
\appendix
%=============================================================================
\section{Condition of detailed balance for the dynamics
(\ref{equations_of_motion})}
\label{detailed-balance-proof}
We prove here that the dynamics defined by the equations of motion
(\ref{equations_of_motion}) satisfies detailed balance if and only if $c_k=c$ for all $k$, that is, if the stochastic forcing has a white spectrum in space. 

We start from the $N$-particle Fokker-Planck equation
(\ref{eq:N-particle-Fokker-Planck}) associated with the equations of motion (\ref{equations_of_motion}). It will be useful to rewrite it in the following way:
\begin{equation}\label{fokker-planck-appendix}
\frac{\p f_N(\bold{x})}{\p t}=-\sum_{i=1}^{2N} \frac{\p}{\p x_i}[A_i(\bold{x}) f_N(\bold{x})] + \frac{1}{2}\sum_{i,j=1}^{2N}\frac{\p^2}{\p x_i\p x_j}[B_{i,j}(\bold{x}) f_N(\bold{x})]\,,
\end{equation}
where $x_i=q_i$ for $i=1,...,N$, $x_i=p_{i-N}$ for $i=(N+1),...,2N$, and
we use the notation $\bold{x}=\{x_i\}$. The drift vector $A_i(\bold{x})$
is a function of the $x_i$'s, and is given by 
\begin{eqnarray}\label{drift-appendix}
A_i(\bold{x})&=p_i\qquad {\rm for } \qquad i=1,...,N,\\
A_i(\bold{x})&=-\alpha p_{i-N}-\frac{1}{N}\sum_{j=1}^N\frac{\p v(q_{i-N}-q_j)}{\p q_{i-N}}\qquad {\rm for } \qquad i=(N+1),...,2N.
\end{eqnarray}
Similarly, the expression for the (symmetric) diffusion matrix $B_{i,j}$ is:
\begin{eqnarray}\label{diffusion-appendix}
B_{i,j}(\bold{x})=\alpha C(|q_{i-N}-q_{j-N}|)\qquad {\rm for}\qquad i>
N\,\wedge\,j> N,
\end{eqnarray}
and $B_{i,j}(\bold{x})=0$ otherwise. We moreover introduce the constants $\epsilon_i=\pm 1$, which denote the parity with respect to time inversion of the variables $x_i$, and the notation $\epsilon \bold{x}=\{\epsilon_i x_i\}$.

It can be shown (see \cite{Gardiner}, Sect. 5.3.5, or \cite{Risken}, Sect. 6.4) that the dynamics described by a Fokker-Planck equation of the form 
(\ref{fokker-planck-appendix}) satisfies detailed balance if and only if
the following two conditions are satisfied ($i=1,...,2N$):
\begin{equation}\label{condition-1-appendix-db}
\epsilon_i\epsilon_jB_{i,j}(\bold{\epsilon x})=B_{i,j}(\bold{x}),
\end{equation}
and
\begin{equation}\label{condition-2-appendix-db}
\epsilon_i A_i(\epsilon\bold{x})
f^s_N(\bold{x})=-A_i(\bold{x})f^s_N(\bold{x})+\sum_{j=1}^{2N}\frac{\p}{\p
x_j}B_{i,j}(\bold{x})f^s_N(\bold{x}),
\end{equation}
where $f^s_N(\bold{x})$ is the stationary solution of (\ref{fokker-planck-appendix}).

In our case, in which the drift and the diffusion terms are given by
Eqs.~(\ref{drift-appendix}) and (\ref{diffusion-appendix}),
respectively, the condition (\ref{condition-1-appendix-db}) is trivially
satisfied. Our proof goes as follows: we solve formally Eq.~(\ref{condition-2-appendix-db}), and show that $f^s_N(\bold{x})$ is a
stationary solution of Eq.~(\ref{fokker-planck-appendix}) if and only if
the non-vanishing part of $B_{i,j}$ is proportional to the identity
matrix. Then, it is simple to show that this implies that the spectrum of the forcing has to be white in space.

Equation~(\ref{condition-2-appendix-db}) for $i=1,...,N$ is also
trivially satisfied. On the other hand, for what concerns $i=(N+1),...,2N$, we have
\begin{equation}\label{condition-2-appendix-db-2}
2p_kf^s_N(\bold{x})=-\sum_{j=1}^N C(|q_k-q_j|)\frac{\p f^s_N(\bold{x})}{\p p_j},
\end{equation}
where $k=i-N$. We introduce the $N\times N$ matrix $\mathcal{C}$ whose
components are given by $\mathcal{C}_{k,j}(\bold{x})=C(|q_k-q_j|)$, and
observe that, for generic values of the $q_i$'s, $\mathcal{C}$ admits an
inverse $\mathcal{C}^{-1}$. Integrating Eq.~(\ref{condition-2-appendix-db-2}), we thus have
\begin{equation}\label{stationary-solution-appendix-db}
f^s_N(\bold{x})=d(q_1,...,q_N)\exp\left[-\sum_{k,j=1}^N p_k \left(\mathcal{C}^{-1}\right)_{k,j}p_j\right],
\end{equation}
where $d(q_1,...,q_N)$ is an undetermined function.
Inserting Eq.~(\ref{stationary-solution-appendix-db}) into the Fokker-Planck equation (\ref{fokker-planck-appendix}), imposing that it is a stationary solution, and with some calculations, we get
\begin{equation}\label{stationary-solution-appendix-db-2}
\sum_{i=1}^N\left[-\frac{\p f^s_N}{\p q_i}\frac{\p H}{\p p_i}+\frac{\p H}{\p q_i}\frac{\p f^s_N}{\p p_i}\right]=0\,.
\end{equation}
Then, $f^s_N$ is a function of the Hamiltonian $H$, that is
$f^s_N(\bold{x})=\psi(H(\bold{x}))$ for some function~$\psi$. On the
other hand, because $f^s_N$ is given by the formula in Eq.~(\ref{stationary-solution-appendix-db}), we can also deduce that $\psi$
is an exponential, and thus, that $f^s_N$ is Gaussian in the velocities.
We conclude that $\mathcal{C}^{-1}$ (and hence, $\mathcal{C}$) has to be independent of the $q_i$'s
and proportional to the identity.
Finally, from the form of $C(|q_i-q_j|)$ in Eq.~(\ref{ckdefinition}), we see that this condition on $\mathcal{C}$ is satisfied if and only if the spectrum of the forcing is white in space.
%=============================================================================
\section{Closure of the BBGKY hierarchy (\ref{BBGKY-stochastic})}\label{Appendix_closure_BBGKY}
\label{BBGKYclosure}
We analyze here in detail the closure of the BBGKY hierarchy discussed in
the text, in particular, the reasons for which the connected part of the
two-particle correlation is of order $\alpha$, while higher correlations
are negligible at leading order in $\alpha$, so that this closure is self-consistent.

In the following, we expand the functions $f_2$ and $f_3$ as
\begin{equation}\label{2-particle-connected-correlations-appendix}
f_2(z_1,z_2,t)=f(z_1,t)f(z_2,t)+ \tilde{g}(z_1,z_2,t),
\end{equation}
and
\begin{eqnarray}\label{3-particle-connected-correlations-appendix}
f_3(z_1,z_2,z_3,t)&=&f(z_1,t)f(z_2,t)f(z_3,t)+f(z_1,t)\tilde{g}(z_2,z_3,t)\nonumber
\\
&&+ f(z_2,t)\tilde{g}(z_1,z_3,t)+ f(z_3,t)\tilde{g}(z_1,z_2,t)+
h(z_1,z_2,z_3,t),
\end{eqnarray}
and similarly, for other $f_s$'s for $s \ge 4$.

Now, let us write explicitly the first two equations of the BBGKY hierarchy
(\ref{BBGKY-stochastic}). The first one, obtained from Eqs.~(\ref{BBGKY-stochastic}) and
(\ref{2-particle-connected-correlations-appendix}), is
\begin{equation}\label{eq:stochastic-BBGKY-1-eq-appendix}
\frac{\partial f}{\partial t}+p\frac{\p f}{\p q}-\frac{\p f}{\p
p}\frac{\p \Phi[f]}{\p q}-\alpha\frac{\partial}{\partial
p}[pf]-\frac{\alpha}{2}C(0)\frac{\partial^{2}f}{\partial
p^{2}}=\frac{\partial}{\partial p}\int{\rm d}q_{1}{\rm
d}p_{1}\,v'(q-q_1)\tilde{g}(z,z_1,t),
\end{equation}
where 
\begin{equation}
\Phi[f](q)=\int {\rm d}q_{1}{\rm d}p_{1}\,v(q-q_1)f(q_1,p_1,t)
\end{equation}
is the mean-field potential. For the second equation of the hierarchy,
we use Eqs.~(\ref{3-particle-connected-correlations-appendix}) and
(\ref{eq:stochastic-BBGKY-1-eq-appendix}) to get
\begin{eqnarray}\label{eq:stochastic-BBGKY-2-eq-appendix}
\frac{\p \tilde{g}(z_1,z_2,t)}{\p t}&=&\Big[- p_1\frac{\p \tilde{g}}{\p
q_1}+\frac{\p \tilde{g}}{\p p_1}\frac{\p \Phi[f]}{\p
q_1}+\frac{f(z_2)}{N}\frac{\p v(q_1-q_2)}{\p q_1}\frac{\p f}{\p
p_1}+\frac{1}{N}\frac{\p v(q_1-q_2)}{\p q_1}\frac{\p \tilde{g}}{\p
p_1}\nonumber\\
&&+\frac{\p f}{\p p_1}\int {\rm d}z_3 \frac{\p v(q_1-q_3)}{\p q_1}\tilde{g}(z_2,z_3)
+\frac{\p}{\p p_1}[\alpha p_1 \tilde{g}]+\frac{\alpha}{2}C(|q_1-q_2|)\frac{\p f}{\p p_1}\frac{\p f}{\p p_2}\nonumber\\
&&+\frac{\alpha}{2}C(0)\frac{\p^2 \tilde{g}}{\p
p_1^2}+\frac{\alpha}{2}C(|q_1-q_2|)\frac{\p^2  \tilde{g}}{\p p_1 \p
p_2}+\int {\rm d}z_3\, \frac{\p v(q_1-q_3)}{\p q_1}\frac{\p h}{\p
p_1}\Big]\nonumber \\
&&+\{1\leftrightarrow 2\},
\end{eqnarray}
where the symbol
$\{1\leftrightarrow 2\}$ stands for an expression obtained from the
bracketed one on the right hand side by exchanging the subscripts $1$ and $2$.

Let us analyze the order of magnitude of various terms in Eq.~(\ref{eq:stochastic-BBGKY-2-eq-appendix}). First of all, we have $f\sim
1$, as it is normalized to unity. However, we do not know a priori the
order of magnitude of $\tilde{g}$ and $h$. Thus, the order of magnitude
of all but the terms $
\frac{f(z_2)}{N}\frac{\p v(q_1-q_2)}{\p q_1}\frac{\p f}{\p p_1}$ and
$\frac{\alpha}{2}C(|q_1-q_2|)\frac{\p f}{\p p_1}\frac{\p
f}{\p p_2}$ is unknown. In the continuum limit $N\alpha\gg 1$, we have
\begin{equation}
\frac{f(z_2)}{N}\frac{\p v(q_1-q_2)}{\p q_1}\frac{\p f}{\p p_1}\sim
\frac{1}{N}\ll \alpha\sim \frac{\alpha}{2}C(|q_1-q_2|)\frac{\p f}{\p
p_1}\frac{\p f}{\p p_2},
\end{equation}
so that it is natural to guess that $\tilde{g}\sim \alpha$. Let us also observe that in the limit $N\alpha\ll 1$, we have
\begin{equation}
\frac{f(z_2)}{N}\frac{\p v(q_1-q_2)}{\p q_1}\frac{\p f}{\p p_1}\sim
\frac{1}{N}\gg \alpha\sim \frac{\alpha}{2}C(|q_1-q_2|)\frac{\p f}{\p
p_1}\frac{\p f}{\p p_2},
\end{equation}
so that we obtain $\tilde{g}\sim 1/N$. In the limit $N\alpha \ll 1$, the kinetic theory leads to the Lenard-Balescu equation.

Once we have established that $\tilde{g}\sim \alpha$, one can write down
the equation of the hierarchy for $h$ and, with similar reasoning as
above, one
then finds that $h$ is at least of order $\alpha/N\ll \alpha$ (or,
$\alpha^2$ depending on whether $ \alpha/N\ll \alpha^2$ or the reverse), so that the term
$\int {\rm d}z_3\, \frac{\p v(q_1-q_3)}{\p q_1}\frac{\p h}{\p p_1}$
is negligible in Eq.~(\ref{eq:stochastic-BBGKY-2-eq-appendix}), as may
be straightforwardly checked. The iterative procedure can be repeated at all orders of the hierarchy.
Discarding three-particle and higher-order correlations is thus a
self-consistent procedure. Moreover, note that in Eq.~(\ref{eq:stochastic-BBGKY-2-eq-appendix}), some of the terms are of
higher orders ($\alpha^2$, $\alpha/N$,...) with respect to $\alpha$, and
thus, can be discarded. The final form of the second equation of the BBGKY hierarchy is thus
\begin{eqnarray}
\frac{\p \tilde{g}(z_1,z_2)}{\p t}&=&\Big[-p_1\frac{\p \tilde{g}}{\p q_1}+\frac{\p \tilde{g}}{\p p_1}\frac{\p \Phi[f]}{\p q_1}
+\frac{\p f}{\p p_1}\int {\rm d}z_3 v'(q_1-q_3)\tilde{g}(z_2,z_3)
\nonumber \\
&&+\frac{\alpha}{2}C(|q_1-q_2|)\frac{\p f}{\p p_1}\frac{\p f}{\p
p_2}\Big]+\{1\leftrightarrow 2\}.
\end{eqnarray}
Note that $\tilde{g}\sim \alpha$ implies, see Eq.~(\ref{eq:stochastic-BBGKY-1-eq-appendix}), that the mean-field effect of
the stochastic forces gives a contribution at the same order to the two-particle correlation induced by them.
%=============================================================================
\section{Evolution of the kinetic energy for the dynamics
(\ref{equations_of_motion})}
\label{kinetic-energy-evolution}
We derive here the evolution of the kinetic energy as obtained from the
kinetic equation (\ref{eq:stochastic-kinetic-equation}). Let us recall
that the average kinetic energy density at time $t$ in the continuous limit can be written as
\begin{equation}
\langle k(t)\rangle =\frac{1}{2}\int dp\, p ^2\, f(p,t).
\end{equation}
The starting point to obtain its time evolution is to multiply the
kinetic equation (\ref{eq:stochastic-kinetic-equation}) by
$\frac{1}{2}p^2$, and then, to integrate over $p$. Neglecting for the moment
the non-linear part of the diffusion coefficient, and integrating by
parts, we get
\begin{equation}\label{kinetic_energy_evolution_diff_eq}
\left \langle \frac{\p k(t)}{\p t}\right\rangle\,+\,2\alpha\,\langle
k(t)\rangle \,-\,\frac{\alpha}{2}C(0)\,=0,
\end{equation}
which gives
\begin{equation}\label{kinetic_energy_evolution}
\langle k(t) \rangle \,=\,\left(\langle k(0)\rangle-\frac{C(0)}{4}\right)\,e^{-2\alpha t}\,+\,\frac{C(0)}{4}.
\end{equation}
The kinetic energy density in the stationary state is thus
$\left\langle \kappa\right\rangle_{ss} = C(0)/4$.

We now have to prove that the non-linear part of the diffusion
coefficient (\ref{diffusion_coefficient}) does not contribute to the
time evolution of the kinetic energy. Such a result is expected and is
usually valid for collisional terms (i.e., those terms in the kinetic
equations which are given by two-particle correlation), for example, in the
Boltzmann equation or in the Lenard-Balescu equation \cite{Balescu:1997}. 
The contribution to $k(t)$ from the non-linear part of the diffusion coefficient is a sum of terms proportional to
\begin{eqnarray}\label{appendix_b_1}
T&=&\frac{1}{2}\int dp \,p ^2\,\nonumber \\
&&\times\frac{\p}{\p p}\,\left\{f'(p,t)
\left[\frac{1}{|\epsilon(k,k
p)|^2}\,\int^*dp_1\,\frac{f'(p_1,t)}{p_1-p}\,+\,\int^*dp_1\,\frac{f'(p_1,t)}{p_1-p}\frac{1}{|\epsilon(k,k
p_1)|^2}\right]\right\}; \nonumber \\
\end{eqnarray}
we will show that each of such terms vanishes independently. Indeed,
integrating the last expression over $p$ by parts, we get that
\begin{equation}
T=-\int dp\int ^* dp_1\,p\,f'(p,t)\left[\frac{f'(p_1,t)}{p_1-p}\,\frac{1}{|\epsilon (k,k p)|^2}+\frac{f'(p_1,t)}{p_1-p}\,\frac{1}{|\epsilon (k,k p_1)|^2}\right].
\end{equation}
Exchanging now the variables $p_1$ and $p$ and the order of integration,
we get that the above equation may be rewritten as
\begin{equation}
T=\int dp\int ^* dp_1\,p_1\,f'(p_1,t)\left[\frac{f'(p,t)}{p_1-p}\,\frac{1}{|\epsilon (k,k p_1)|^2}+\frac{f'(p,t)}{p_1-p}\,\frac{1}{|\epsilon (k,k p)|^2}\right]\,.
\end{equation}
Summing up the last two equations, we therefore have 
\begin{equation}
T=\frac{1}{2}\int dp \,\int dp_1\,f'(p_1,t)\,f'(p,t)\left[\frac{1}{|\epsilon (k,k p_1)|^2}+\frac{1}{|\epsilon (k,k p)|^2}\right],
\end{equation}
which vanishes on integrating by parts both with respect to $p_1$ and $p$.
%=============================================================================
\section{Proof that Eq.~(\ref{eq:stochastic-kinetic-equation}) admits non-Gaussian stationary
distribution with Gaussian tails}
\label{nonGaussian}
We prove here that for a general forcing spectra, the Gaussian
distribution function in Eq.~(\ref{Gaussian}) is not a stationary
solution of the kinetic equation (\ref{eq:stochastic-kinetic-equation}), and that the tails of any stationary state are Gaussian.
For the first point, we have to prove that the contribution to $\partial
f/\partial t$ from the non-linear part of $D[f](p)$ in Eq.~(\ref{diffusion_coefficient}) is not vanishing. This result can be
proven with an asymptotic expansion \cite{Lifshitz:2002} for large momenta of the integrals which appear in the diffusion coefficient. Given any function $g(p)$, we approximate integrals of the form 
\begin{equation}
\int^* dp_1 \frac{g(p_1)}{p_1-p}
\end{equation}
by expanding $\frac{1}{p_1-p}$ in Taylor series. We get, for example,
\begin{equation}
\int^* {\rm d}p_1\, \frac{f'_G(p_1)}{p_1-p}\,\simeq\,\frac{2}{\sqrt{\pi}}\beta^{3/2}\int_{-\infty}^{\infty} e^{-\beta p_1^2}\left[\frac{p_1}{p}+\left(\frac{p_1}{p}\right)^2+\left(\frac{p_1}{p}\right)^3+...\right]\,\simeq\,\frac{1}{p^2},
\end{equation}
where, in the last equality, we have taken into account the fact that
the Gaussian distribution being even, the terms containing
$\left(\frac{p_1}{p}\right)^k$ with $k$ odd do not contribute. In a
similar way, we have
\begin{equation}
|\epsilon(k,k p)|^2\,\simeq\,1-\frac{4\pi v(k)}{p^2},
\end{equation}
and
\begin{equation}
\int^* dp_1\,\left[\frac{f'_G(p_1)}{p_1-p}\frac{1}{|\epsilon(k,k p_1)|^2}\right]\,\simeq\,\frac{2\beta ^{3/2}}{\sqrt{\pi}\,p^2}\,\int dp_1\,\frac{p_1^2\,e^{-\beta p_1^2}}{|\epsilon(k,k p_1)|^2},
\end{equation}
where we have used the fact that $|\epsilon(k,k p)|^2$ is an even
function of $p$. With these results, we can evaluate the non-linear part
of the kinetic equation: for large $p_1$, the non-linear contribution to $\p f/\p t$ is
\begin{equation}\label{stat_state_non_gaussian}
2\pi\alpha\sum_{k=1}^{\infty}\,v_k\,c_k\left[1+\frac{2\beta^{3/2}}{\sqrt{\pi}}\,\int \,dp\,\frac{p^2 e^{- p^2}}{|\epsilon(k,k p)|^2}\right]\left[\frac{4\beta^{5/2}}{\sqrt{\pi}}\,e^{-\beta p_1^2}\right].
\end{equation}
It can be shown that such a term is a non-vanishing function of
$p_1$. This completes the proof: for a generic forcing spectra, the
stationary state, when exists, is not Gaussian. 

Using the same asymptotic expansion as before, it can be checked
that the diffusion coefficient $D[f](p)$ converges to $C(0)/2$ for
any distribution $f$. From this observation and Eq.~(\ref{stationary-FokkerPlanck}), it follows that any stationary solution
of the kinetic equation~(\ref{eq:stochastic-kinetic-equation}) has Gaussian tails.
%=============================================================================


\section*{References}
\begin{thebibliography}{10}
\bibitem{Campa:2009}Campa A, Dauxois T and Ruffo S, {\em Statistical
mechanics and dynamics of solvable models with long-range interactions},
2009 {\em Phys. Rep.} {\bf 480} 57
\bibitem{Bouchet:2010}Bouchet F, Gupta S and Mukamel D, {\em Thermodynamics and dynamics of systems with long-range interactions}, 2010 {\em
Physica A} {\bf 389} 4389
\bibitem{Chavanis}Chavanis P H, {\em Phase transitions in self-gravitating systems}, 2006 {\em Int. J. Mod. Phys. B} {\bf 20} 3113
\bibitem{JStatMech}{\em J. Stat. Mech.} Topical Issue: Long-Range
Interacting Systems 
\bibitem{Bouchet-Venaille}Bouchet F and Venaille A, {\em Statistical
mechanics of two-dimensional and geophysical flows}, 2012 {\em Phys.
Rep.} {\bf 515} 227.
\bibitem{Weinberg2000}Weinberg M D, {\em Noise driven evolution in stellar systems I. Theory}, 2001 {\em Mon. Not. R. Astron. Soc.} {\bf 328} 311
\bibitem{Liewer}Liewer P C, {\em Measurements of microturbulence in tokamaks and comparisons with theories of turbulence and anomalous transport}, 1985 {\em Nucl. Fusion} {\bf 25} 543
\bibitem{Dhar:2008}Dhar A, {\em Heat transport in low-dimensional systems}, 2008 {\em Adv. Phys.} {\bf 57} 457
\bibitem{Derrida:2007}Derrida B, {\em Non-equilibrium steady states: Fluctuations and large deviations of the density and of the current}, 2007 {\em J. Stat. Mech.} P07023
\bibitem{Jarzynski:2008}Jarzynski C, {\em Nonequilibrium work relations:
Foundations and applications}, 2008 {\em Eur. Phys. J. B} {\bf 64} 331
\bibitem{Nicholson:1992}Nicholson D R, {\em Introduction to plasma
physics}, 1992 (Krieger, Malabar, Florida)
\bibitem{Lifshitz:2002}Lifshitz E M and Pitaevski L P, {\em Physical
kinetics}, 2002 (Butterworth-Heinemann, London)
\bibitem{Bouchet-Simonnet}Bouchet F and Simonnet E, {\em Random changes of flow topology in two-dimensional and geophysical turbulence}, 2009 {\em Phys. Rev. Lett.} {\bf 102} 94504
\bibitem{Briggs}Briggs R J, Daugherty J D and Levy R H, {\em Role of
Landau damping in crossed-field electron beams and inviscid shear flow},
1970 {\em Phys. Fluids} {\bf 13} 421
\bibitem{Dikii}Dikii L A, {\em The stability of plane-parallel flows of
an ideal fluid}, 1960 {\em Dokl. Akad. Nauk SSSR} {\bf 135} 1068,
Translated in {\em Sov. Phys. Doklady} {\bf 5}, 1179 
\bibitem{Nardini:2012}Nardini C, Gupta S, Ruffo S, Dauxois T and Bouchet
F, {\em Kinetic theory for non-equilibrium stationary states in long-range interacting systems}, 2012 {\em J. Stat. Mech.} L01002
\bibitem{Chavanis:2012}Chavanis P H, {\em Kinetic theory of spatially
homogeneous systems with long-range interactions: I. General results},
2012 {\em Eur. Phys. J PLUS} {\bf  127} 19 
\bibitem{Milion-body-pb}Heggie D and Hut P, {\em The gravitational million-body problem}, 2003 (Cambridge University Press, Cambridge, UK)
\bibitem{Kac}Kac M, Uhlenbeck G E and Hemmer P C, {\em On the van der Waals theory of the vapor-liquid equilibrium I. Discussion of a one-dimensional Model}, 1963 {\em J. Math. Phys.} {\bf 4} 216
\bibitem{Papoulis}Papoulis A, {\em Probability, random variables and stochastic
processes}, 1965 (Tokyo: McGraw-Hill Kogakusha)
\bibitem{Bellet}Rey-Bellet L, {\em Ergodic properties of Markov
processes}, in {\em Open Quantum systems II. The Markovian approach,
Lecture notes in Mathematics 1881}, 2006 (Springer, Berlin)
\bibitem{Gardiner}Gardiner C W, {\em Handbook of stochastic methods
for physics, chemistry and the natural sciences}, 1983 (Springer-Verlag, Berlin)
\bibitem{Antoni}Antoni M and Ruffo S, {\em Clustering and relaxation in Hamiltonian long-range dynamics}, 1995 {\em Phys. Rev. E} {\bf 52} 2361
\bibitem{yamaguchi2004}Yamaguchi Y Y, Barr{\'e} J, Bouchet F, Dauxois T and Ruffo S, {\em Stability criteria of the Vlasov equation and quasi-stationary states of the HMF model}, 2004 {\em Physica A} {\bf 337} 36
\bibitem{Risken}Risken H, {\em The Fokker-Planck equation: Methods of solutions and applications}, 1989 (Springer-Verlag, Berlin)
\bibitem{Huang}Huang K, {\em Statistical mechanics}, 1987 (Wiley, New York)
\bibitem{Nardini-Bouchet}Nardini C and Bouchet F, 2013 (in preparation)
\bibitem{Balescu:1997}Balescu R, {\em Statistical Dynamics: Matter Out
of Equilibrium}, 1987 (Imperial College Press, London) 
\bibitem{McLachlan}McLachlan R I and Atela P, 1992 {\em Nonlinearity}
{\bf 5} 541 
\bibitem{Baldovin_Orlandini}Baldovin F and Orlandini E, {\em
Nos\'e-Hoover and Langevin thermostats do not reproduce the nonequilibrium behavior of long-range Hamiltonians}, 2007 {\em Int. J. Mod. Phys. B} {\bf 21} 4000
\bibitem{ChavanisBMF}Chavanis P H, Baldovin F and Orlandini E, {\em
Noise-induced dynamical phase transitions in long-range systems}, 2011 {\em
Phys. Rev. E} {\bf 83} 040101(R)
\bibitem{Frank}Frank T D, {\em Nonlinear Fokker-Planck Equations:
Fundamentals and Applications}, 2005 (Springer, Berlin)
\bibitem{Arnold}Arnold L, {\em Random dynamical systems}, 2003
(Springer-Verlag, Berlin) 
\bibitem{Maritan-etal}Maritan A and Banavar J R, {\em Chaos, noise, and
synchronization}, 1994 {\rm Phys. Rev. Lett.} {\bf 72} 1451; Pikovsky A
S, {\em Comment on ``Chaos, Noise, and Synchronization"} 1994 {\em Phys.
Rev. Lett.} {\bf 73} 2931; Maritan A and Banavar J R, {\em  Maritan and
Banavar Reply} 1994 {\em Phys. Rev. Lett.} {\bf 73} 2932 
\bibitem{Kramers}Kramers H A, {\em Brownian motion in a field of force
and the diffusion model of chemical kinetics}, 1940 {\rm Physica} {\bf 7} 284
\bibitem{vanKampen}van Kampen N G, {\em Stochastic processes in physics
and chemistry}, 1992 (North Holland, Amsterdam)
\bibitem{Hanggi}H\"{a}nggi P, Talkner P and Borkovec M, {\em Reaction
Rate Theory: Fifty Years After Kramers} 1990 {\em Rev. Mod. Phys.} {\bf
62} 251
\bibitem{Freidlin}Freidlin M I and Wentzell A D {\em Random
perturbations of dynamical systems}, 1998 (Springer, Berlin)
\bibitem{Hugo}Touchette H, {\em The large deviation approach to
statistical mechanics}, 2009 {\em Phys. Rep.} {\bf 478} 1 
\bibitem{Heyvaerts}Heyvaerts J, {\em A Balescu-Lenard-type kinetic
equation for the collisional evolution of stable self-gravitating systems}, 2010 {\em Mon. Not. R. Astron. Soc.} {\bf 407} 355
\end{thebibliography}
\end{document}